\documentclass[twocolumn, prl,superscriptaddress,
amsmath,amssymb, 
aps,floatfix,]{revtex4-2}

\usepackage{mathtools}
\usepackage{graphicx}
\usepackage{dcolumn}
\usepackage{bm}
\usepackage{bbold} 
\usepackage{hyperref}
\usepackage[dvipsnames]{xcolor}
\renewcommand{\selectlanguage}[1]{}

\begin{document}

\title{Canard cascading in networks with adaptive mean-field coupling} 

\author{J.~Balzer}
\affiliation{Institut für Theoretische Physik, Technische Universität Berlin,
Hardenbergstraße 36, 10623 Berlin, Germany}

\author{R.~Berner}
\affiliation{Department of Physics, Humboldt-Universität zu Berlin, Newtonstr. 15, Berlin, 12489, Germany}

\author{K.~Lüdge}
\affiliation{Technische Universität Ilmenau, Institut für Physik, Weimarer Straße 25, 98693 Ilmenau, Germany}

\author{S.~Wieczorek}
\affiliation{School of Mathematical Sciences, University College Cork, Ireland}

\author{J.~Kurths}
\affiliation{Department of Physics, Humboldt-Universität zu Berlin, Newtonstr. 15, Berlin, 12489, Germany}
\affiliation{Potsdam Institute for Climate Impact Research (PIK), Potsdam, Germany}

\author{S.~Yanchuk}
\affiliation{School of Mathematical Sciences, University College Cork, Ireland}
\affiliation{Potsdam Institute for Climate Impact Research (PIK), Potsdam, Germany}

\date{\today}

\begin{abstract}
Canard cascading (CC) is observed in dynamical networks with global adaptive coupling. It is a fast-slow phenomenon characterized by a recurrent sequence of fast transitions between distinct and slowly evolving quasi-stationary states.
In this letter, we uncover the dynamical mechanisms behind CC, using an illustrative example of globally and adaptively coupled semiconductor lasers, where CC represents sequential switching on and off the lasers.
Firstly, we show that CC is a robust and truly adaptive network effect that is scalable with  network size and does not occur without adaptation.
%
Secondly, we 
 uncover multiple saddle slow manifolds (unstable quasi-stationary states) linked by 
heteroclinic orbits (fast transitions) in the phase space of the system. This allows us to identify CC with a novel {\em heteroclinic canard orbit} that organises different unstable quasi-stationary states into an intricate fast-slow limit cycle. Although individual quasi-stationary states are unstable (saddles), the CC cycle as a whole is attractive and robust to parameter changes.
\end{abstract}

\maketitle

%
{\em Dynamical networks} with {\em dynamic nodes} and {\em static links} are famously universal mathematical models used to describe  challenging real-world applications, such as coupled optoelectronic devices, neural networks or power grids \cite{Pikovsky2001,Boccaletti2006,Soriano2013,Gerstner2014,Hellmann2018,Schafer2018,Newman2010a,nicosia_collective_2017}. In addition to the complex network structure, their nodes are often fast-slow, meaning that they evolve on multiple time scales \cite{Desroches2012,Kuehn2015}. Typical examples are coupled semiconductor lasers, where the photon lifetime is much shorter than the carrier lifetime \cite{Erneux2019,roosStabilizingNanolasersPolarization2021}, or coupled neurons, where the voltage changes faster than the gating variables \cite{Izhikevich2000}.

%
%
%
%
%
{\em Adaptive dynamical networks} (ADNs) with 
{\em dynamic nodes} and
{\em dynamic links} are even more advanced mathematical models, where the links between the nodes evolve over time depending on the states of the nodes \cite{gross2008adaptive,Berner2023,Sawicki2023PerspectivesSystems}. 
ADNs capture
the interaction between the network  function (changing states of the nodes) and network structure (changing strength and arrangement of the links).
ADNs are essential for different areas of science, for example, neural plasticity \cite{Gerstner2014}, power grid dynamics \cite{Berner2021c}, or decision-making \cite{Chen2016}, to name a few. Furthermore, they exhibit rich and diverse dynamical behaviors, such as frequency clusters \cite{Berner2019}, recurrent synchronization \cite{Thiele2023}, different phase transitions \cite{Fialkowski2023}, explosive synchronization \cite{zhou2006dynamical,AVA18}. However,  analysis of ADNs is
more demanding 
and often requires new approaches \cite{Berner2023}. 

%
In this letter, we describe a phenomenon of {\em canard cascading} (CC) that emerges from an interplay between bi-stable dynamics of the network nodes and slowly evolving global coupling between the nodes. CC manifests itself as a recurrent sequence of fast transitions between distinct quasi-stationary states.  To the best of our knowledge, a first numerical and experimental report on a variant of this phenomenon was given in \cite{DHuys2021}, in the context of the resonance between the dispersion of the network nodes and the noise strength. Here, we show that CC is a genuine 
ADN phenomenon that does not occur when the coupling is static. We also show that CC is scalable: the number of quasi-stationary states in the sequence increases proportionally to the network size. Crucially, we uncover the  dynamical mechanism underlying CC: a novel heteroclinic canard orbit that organises different unstable quasi-stationary states into an intricate fast-slow limit cycle that is stable and robust to changes in the system parameters.

\begin{figure}
    \centering
    \includegraphics[width=0.8\linewidth]{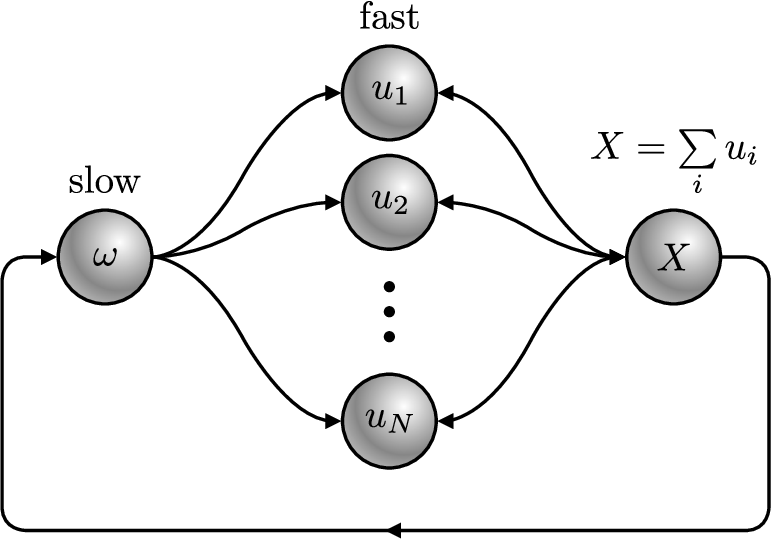}
    \caption{Adaptive dynamical network with global coupling \eqref{eq:general}.  The nodes $u_i$ are globally coupled via two components: the direct mean-field component $X(t)$ and the slowly-adapting component $\omega(t)$.    }
    \label{fig:system_illustration}
\end{figure}

The general theoretical framework for CC has the form
\begin{equation}
\label{eq:general}
\begin{split}
du_i/dt &= g(u_i,\omega , X),\\
    d\omega/dt & = - \varepsilon \, [\omega + f(X)], 
\end{split}
\end{equation}
where $u_i(t)$ is the state of node $i=1,\dots,N$, $X(t) = N^{-1} \sum_{i=1}^N u_i(t)$ is the mean-field, and $\omega(t)$ is the adaptive coupling variable. 
The nodes $u_i(t)$ are globally coupled via 
the direct mean-field $X(t)$ and the adaptive variable $\omega(t)$; see Fig.~\ref{fig:system_illustration}.
The small parameter $\varepsilon\ll 1$ quantifies the ratio of the timescales of the slowly-evolving adaptive coupling $\omega(t)$ and fast-changing nodes $u_i(t)$.


\begin{figure}
    \includegraphics[width=\linewidth]{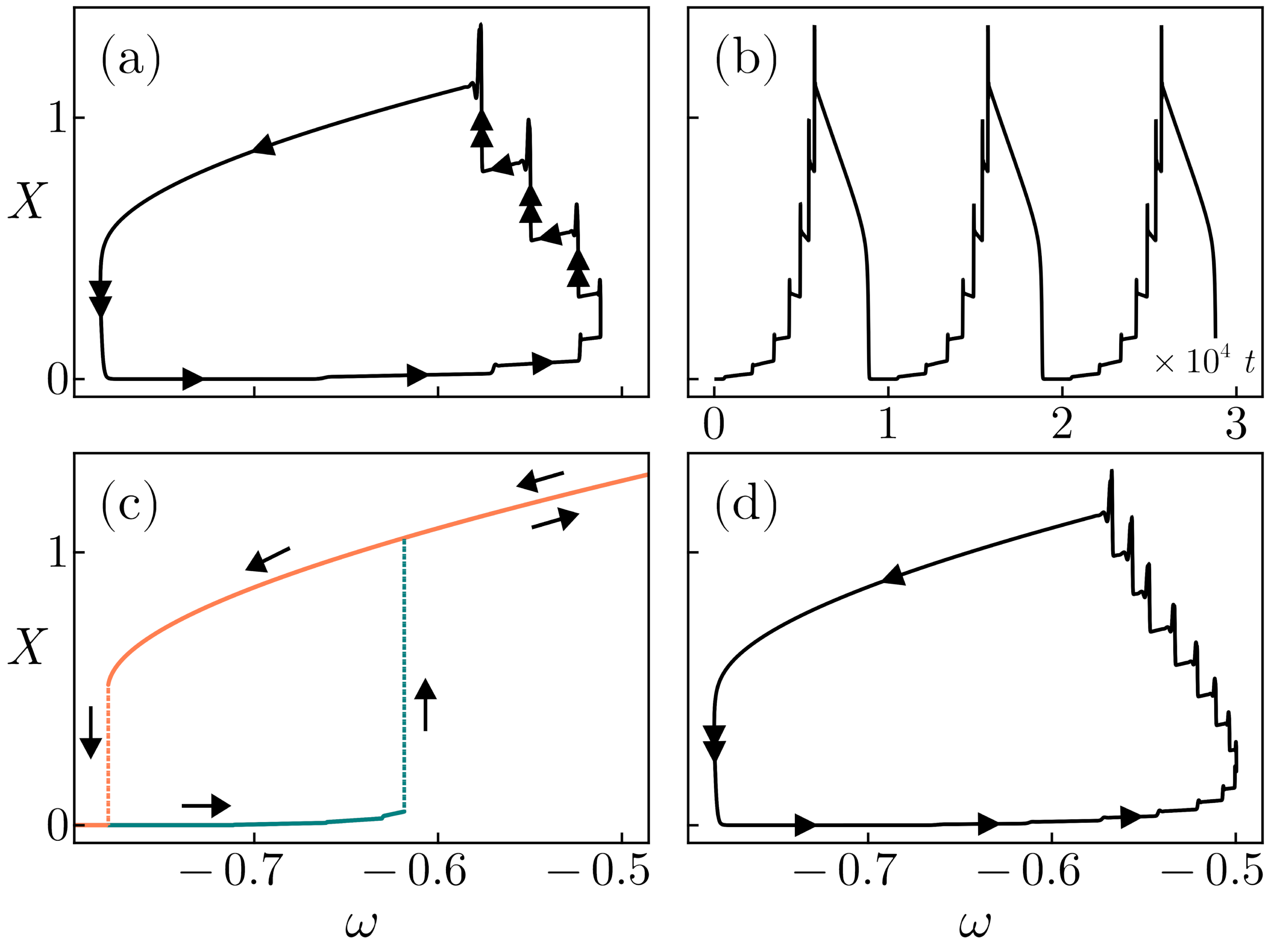}
    \caption{Simulated CC dynamics of system \eqref{eq:laser} with $N=7$ (a-b) and $N=15$ (d) coupled lasers. (a,d) Projection of the solution onto the ($\omega$, $X$)-plane. (b) Time series of the mean-field $X(t)$. 
    (c) The non-adaptive case, where $\omega$ is a parameter, and  parameter scans are performed. The non-adaptive case features a standard hysteresis and no CC. 
    Parameter values: $\gamma = 4 \cdot 10^{-2}$, $\varepsilon = 10^{-4}$,  $\delta_i, i=1, \ldots, N$ are chosen equidistantly in the interval $[1.2, 1.505]$, $k=0.7$, $A=k^{-1}$, and $\alpha=2$.}
    \label{fig:1illustration}
\end{figure}

To demonstrate and analyse CC in a real-life application, we consider a coupled laser model as in \cite{DHuys2021}. Recently, there has been much interest in networks of adaptively coupled optoelectronic devices, owing to their potential for neuromorphic computing \cite{Appeltant2011,Romeira2016,larger_high-speed_2017,Brunner2018,Argyris2020,robertson_ultrafast_2020,Stelzer2021a,apostelReservoirComputingUsing2021,huang_prospects_2022,lupoDeepPhotonicReservoir2023,biasiPhotonicNeuralNetworks2024,henaffOpticalPhaseEncoding2024,deligiannidisMultichannelNonlinearEqualization2024}. Therefore, comprehending the dynamics of such systems, and identifying novel dynamical phenomena, is of importance to future machine learning solutions.
Our  specific model is the network of $N$  coupled semiconductor lasers \cite{dolcemascolo_effective_2020,al-naimee_chaotic_2009}
\begin{align}
dx_i/dt & = x_i  (y_i - 1),  \nonumber \\
dy_i/dt & = \gamma \left[\delta_i - y_i + k \, (\omega + f(X) ) - x_i y_i \right], \label{eq:laser} \\
d\omega/dt & = - \varepsilon \, [\omega + f(X)], \nonumber
\end{align}
where  $x_i(t)$ is the light intensity and $y_i(t)$ is the normalized 
carrier (electron-hole pair) density in laser $i$. The variable $\omega(t)$ is the feedback electric current that plays the role of global nonlinear  adaptive coupling. 
Its evolution is governed by the signal from the
nonlinear amplifier $f(X) = A \ln(1 + \alpha X)$  that receives the mean light intensity $X(t) = N^{-1} \sum_i x_i(t)$, where $A$ and $\alpha$ are two positive feedback parameters. 
Additional parameters include 
electric pump currents $\delta_i$ 
and the 
photodetector responsitivity $k$. 
Our focus will be on a network of non-identical lasers with different $\delta_i$.
CC in the adaptive laser network manifests itself as a sequence of fast `jumps'  in the mean light intensity $X$ (double arrows), each followed by damped oscillations towards a `plateau'  of slowly changing  feedback current $\omega$ with little variation in $X$ (single arrows) in Fig.~\ref{fig:1illustration}(a)-(b). We will show that these slow plateaus correspond to slow motion along an unstable quasi-stationary state (saddle slow manifold). 
Such unusual solutions are known in the literature as {\em canards} 
\cite{benoitChasseAuCanard1981, boldForcedVanPol2003, eckhausRelaxationOscillationsIncluding1983, krupaRelaxationOscillationCanard2001, pengFalseBifurcationsChemical1991, szmolyanCanardsR32001, Wechselberger2005, Wechselberger2013,OSullivan2023}.
Hence the name \textit{canard cascading} (CC). 
The classical examples of low-dimensional fast-slow limit cycles  
have one canard segment and are not robust~\cite{wechselbergerCanards2007}. CC is different in that it  consists of {\em multiple canard segments} and is 
{\em robust}.
For physically meaningful initial conditions and for a wide parameter range, 
the system converges to the {\em CC limit cycle}. 
While a robust non-classical 
slow-fast cycle
with one canard segment has been identified in~\cite{Kuehn2015}, CC appears to be the first example with multiple canard segments.

%


Most importantly, CC is a genuine adaptive network phenomenon: it scales with the network size $N$ and disappears in the absence of adaptation.
When we set $\varepsilon=0$ in 
\eqref{eq:laser}, treat $\omega$ as another parameter, 
a quasi-static sweep in $\omega$ 
uncovers classical hysteresis
in Fig.~\ref{fig:1illustration}(c). One might expect that when $\varepsilon$ is small but non-zero, 
the adaptive network will closely trace out this hysteresis. However, that is not what happens. Instead, an intricate CC emerges during the transition from the lower to the upper part of the hysteresis.
For $N=7$ in Fig.~\ref{fig:1illustration}(a)-(b), this CC consists of seven jumps in $X$ and seven slow plateaus.  Numerical results for different $N$ confirm that the phenomenon  scales with $N$ and appears to be universal, see e.g. the case $N=15$.

We will now describe the mechanism behind CC and show that the fast jumps occur along special heteroclinic orbits connecting different unstable 
quasi-stationary states. We begin by identifying 
quasi-stationary states and their stability. 
In the terminology of geometric singular perturbation theory \cite{Fenichel1979, Jardon-Kojakhmetov2019, Jones1995a, Wechselberger2020}, we identify branches of the critical manifold for system~\eqref{eq:laser}.

To obtain all quasi-stationary states, stable and unstable, we perform the adiabatic elimination of the fast laser field $x_i$ and carrier density $y_i$ by setting $\dot x_i=0$ and $\dot y_i=0$ in system \eqref{eq:laser}.
This gives the following solutions: the $i$-th laser is either ``off"
$(x_i, y_i) = (0 , \delta_i + k [\omega + f(X)])$ or 
``on" 
$(x_i, y_i) = (\delta_i - 1 + k [\omega + f(X)] , 1)$ for all $i$.
In other words, there is one branch of the critical manifold for every combination where some lasers are  
``on" and the other lasers are ``off". If the set of all $N_+$ lasers that are ``on" is denoted with $I_+$,
then the corresponding branch 
of the critical manifold is given by
\begin{equation}
\label{eq: first representation critical manifold}
(x_{i}, y_{i}) = \begin{cases}
(0, \, \delta_i + k [\omega + f(X)]) &\text{ for } i \notin I_+, \\
(\delta_i - 1 + k [\omega + f(X)], \, 1) &\text{ for } i \in I_+,
\end{cases}
\end{equation}
where the mean-field satisfies the self-consistency equation
\begin{align}
X = \frac{N_{+}}{N} \left[ \bar{\delta} - 1 + k(\omega + f(X))\right],
\label{eq: X_I implicit definition}
\end{align}
with $\bar{\delta} = N_+^{-1} \sum_{i \in I_+} \delta_i$ being the average pump of all lasers that are ``on". Since there are $2^N$ different combinations, there are also $2^N$ different branches of the critical manifold leading to $2^N$ possible quasi-stationary states.

\begin{figure}
    \includegraphics[width=\linewidth]{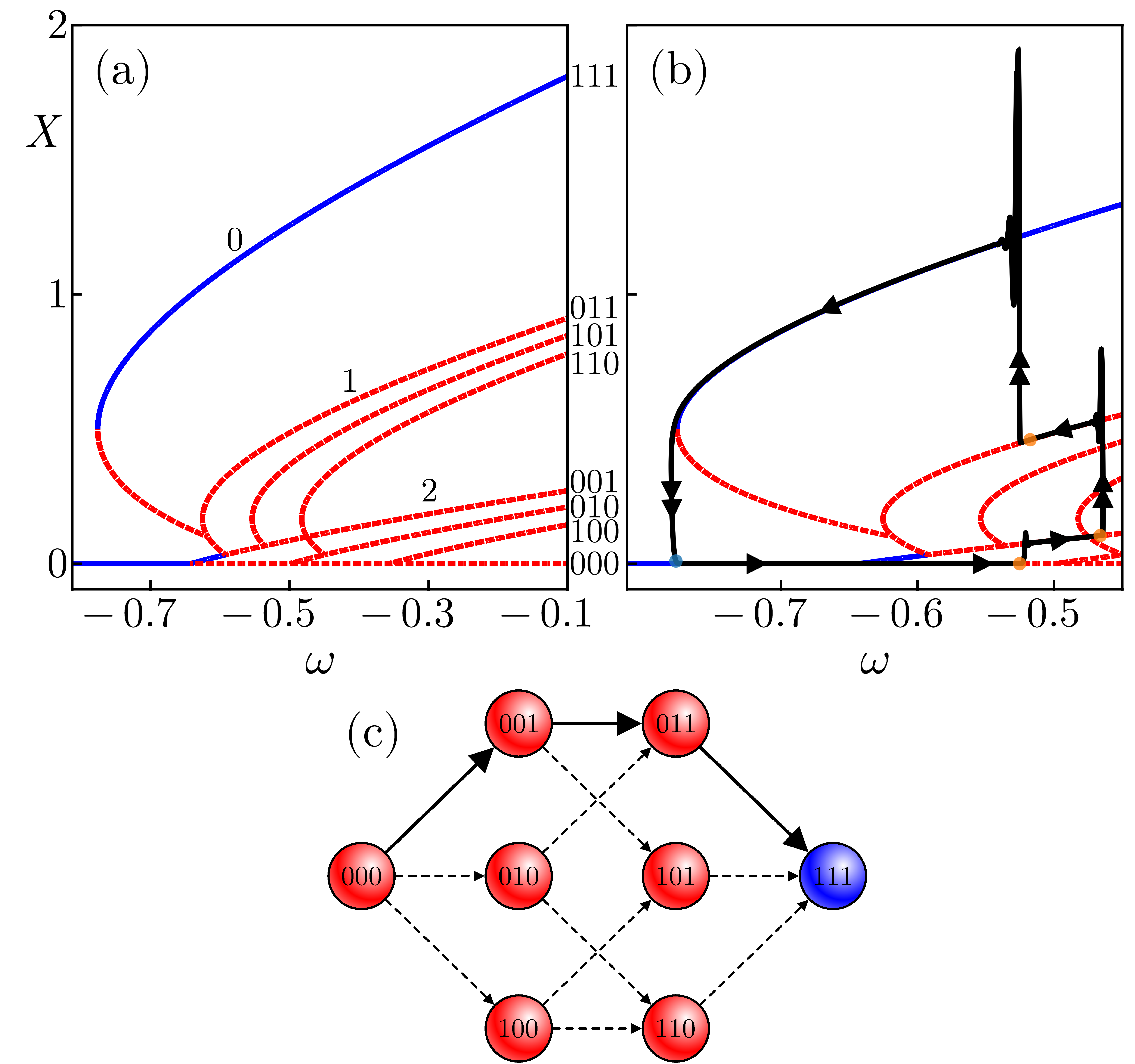}
    \caption{(a) Critical manifold branches of system \eqref{eq:laser} of three coupled lasers.  Projection onto the ($\omega$, $X$)-plane. Stable parts are shown as solid blue lines and unstable parts as dashed red lines. The labels on the right vertical axis show which lasers are "on" (1) or "off" (0). 
    The  numbers on the branches indicate the number of their unstable dimensions.  
    (b) A zoom of part of (a) with CC solution shown by the solid black line. Theoretical predictions for the jump points are shown as orange dots. (c) Schematic representation of the network of heteroclinic connections with solid lines showing the connections along the CC solution. 
    Parameters: $\delta=(1.25, 1.35, 1.45)$, $k=0.7$, $A=k^{-1}$, and $\alpha=2$.
    }
    \label{fig:cm_comparison}
\end{figure}

For $N=3$ lasers, Fig. \ref{fig:cm_comparison}(a) 
shows all $2^N=8$  branches of the critical manifold, in projection onto the $(X,\omega)$ plane. 
Each branch is a  one-dimensional 
curve in the $2N+1=7$-dimensional phase space of 
system \eqref{eq:laser}. 
The stable branches are plotted in blue, while (unstable) saddle branches are plotted in red;
see \cite{supplem} for the 
stability analysis.
We also introduce the notations $B_{000}$, $B_{001}$, ..., $B_{111}$ for the branches of the critical manifold, where ``1" stands for  laser ``on" and ``0" for laser ``off", with the lasers ordered in ascending order of their pump currents $\delta_i$.

The critical manifold has the following properties. 
First, most of the branches are unstable and saddle-type, meaning that they have both repelling and attracting directions. The exceptions are 
$B_{111}$, a part of $B_{000}$, and a small part of $B_{001}$, which are stable.
Two different stable branches co-exist for some interval of $\omega$ in Fig.~\ref{fig:1illustration}(c),  giving rise to bistability and the classical hysteresis upon quasi-static change in $\omega$.
Second, the branches corresponding to the same number $N_{+}$ of active lasers are closely grouped. Their separation 
is proportional to the mismatch in the pump currents $\delta_i$.

The CC limit cycle together with all branches of  the critical manifold 
is shown in Fig.~\ref{fig:cm_comparison}(b). The cycle follows certain branches for significant periods of time, including those identified as unstable (saddles). 
In addition, fast transitions to unstable 
branches are observed. This unusual behaviour raises the following questions.
(q1) Why does the system jump towards unstable branches? (q2) What is the physical interpretation of such jumps?
(q3) How does the system choose a particular unstable branch from each family to jump to? 
(q4)  Why does the system follow these unstable branches for a significant amount of time? 
(q5) Can we estimate this time? 

Owing to the letter format of this publication, we now give conceptual answers to these questions, and move the supporting technical details to \cite{supplem}.

(q1) We begin by answering question (q1). The jumps between two unstable (saddle) branches of the critical manifold are enabled by robust paths between these branches in the phase space, also known as {\em robust heteroclinic connecting orbits} \cite{Krupa1997,musokeSurfaceHeteroclinicConnections2020};
see the schematic diagram in Fig.~\ref{fig:cm_comparison}(c).
The main reasons for the appearance of these heteroclinic orbits are twofold:

Firstly, we observe that branches with more active lasers branch off branches with less active lasers,
e.g. $B_{001}$, $B_{010}$ and $B_{001}$ branch off
$B_{000}$. The branching rule is that if a branch $B_{mnl}$ branches off $B_{ijk}$, then $(m+n+l)-(i+j+k)=1$~\footnote{This rule can be violated in the presence of a symmetry, e.g., when the lasers are identical \cite{supplem}.}.
Such a branching rule leads to the connectivity graph $G$ 
in Fig.~\ref{fig:cm_comparison}(c) for sufficiently large $\omega$.  
For smaller $\omega$, a subgraph of $G$ is realised. 
Each branching point corresponds to a transcritical bifurcation of equilibria in the ``layer system" parametrised by $\omega$ (i.e. system~\eqref{eq:laser} with $\varepsilon=0$). Therefore, for a fixed $\omega$,  each arrow in the graph 
corresponds to a heteroclinic orbit 
in the layer system connecting 
two branches.


%

Secondly, the heteroclinic connections are robust.
This robustness is due to the existence of invariant subspaces. 
For example, for a given $\omega$, the heteroclinic orbit 
$B_{000}\to B_{001}$ lies in the 4-dimensional invariant subspace $x_1=x_2=0$. 
Within the invariant subspace, 
it connects a saddle with one unstable direction to an attractor. The same is true for every other heteroclinic connection in Fig.~\ref{fig:cm_comparison}(c). 
In other words, these heteroclinic connections are robust for the flow restricted to their corresponding invariant subspace, similar to  
\cite{Krupa1997,Ashwin2005,Ashwin2008,Ashwin2010}.
Since changes in $\omega$  preserve invariant subspaces, these heteroclinic orbits are robust to changes in $\omega$. This is why they manifest in the full system with changing $\omega$ (i.e. system~\eqref{eq:laser} with $\varepsilon > 0$). 
More generally, even in the absence of invariant subspaces, these heteroclinic connections would be robust 
because they are transverse intersections of unstable and stable invariant manifolds of two saddle branches, as is also the case in the Olsen model~\cite{musokeSurfaceHeteroclinicConnections2020}, see  more details in \cite{supplem}.

(q2) We have shown that there is a correspondence between the directed graph $G$ in Fig.~\ref{fig:cm_comparison}(c) and the heteroclinic connections among the branches of the critical manifold. These connections have a clear physical meaning: each connection $B_I\to B_J$ with $I=a_1\dots a_i\dots a_N$ and $I=a_1\dots a_i+1\dots a_N$ corresponds to a fast increase of the laser $i$ intensity from 0 to the value $\delta_i - 1 + k [\omega + f(X)]$. 

(q3) According to the obtained graph of heteroclinic connections, the system can potentially evolve along any directed path in this graph during the time evolution starting from the off-state.  However, CC is realised along a specific path in the graph, corresponding to the sequential activation of the lasers with the highest pump current $\delta_i$; see the highlighted path in Fig.~\ref{fig:cm_comparison}(c). 
The theoretical reason for this path selection is
that the highlighted path corresponds to the 
most unstable direction that the system chooses in the case of several "exit possibilities". 
As a consequence, the CC phenomenon does not exist for identical coupled lasers, when 
all $N$ connections from $B_{0\dots 0}$ to the branches with a single laser "on" are equally unstable, and there is no preferred direction that would define a cascading path. As a result, the symmetric system jumps directly to the stable branch $B_{1\dots1}$, avoiding the heteroclinic connections even  though they are present. 


\begin{figure}
    \includegraphics[width=\linewidth]{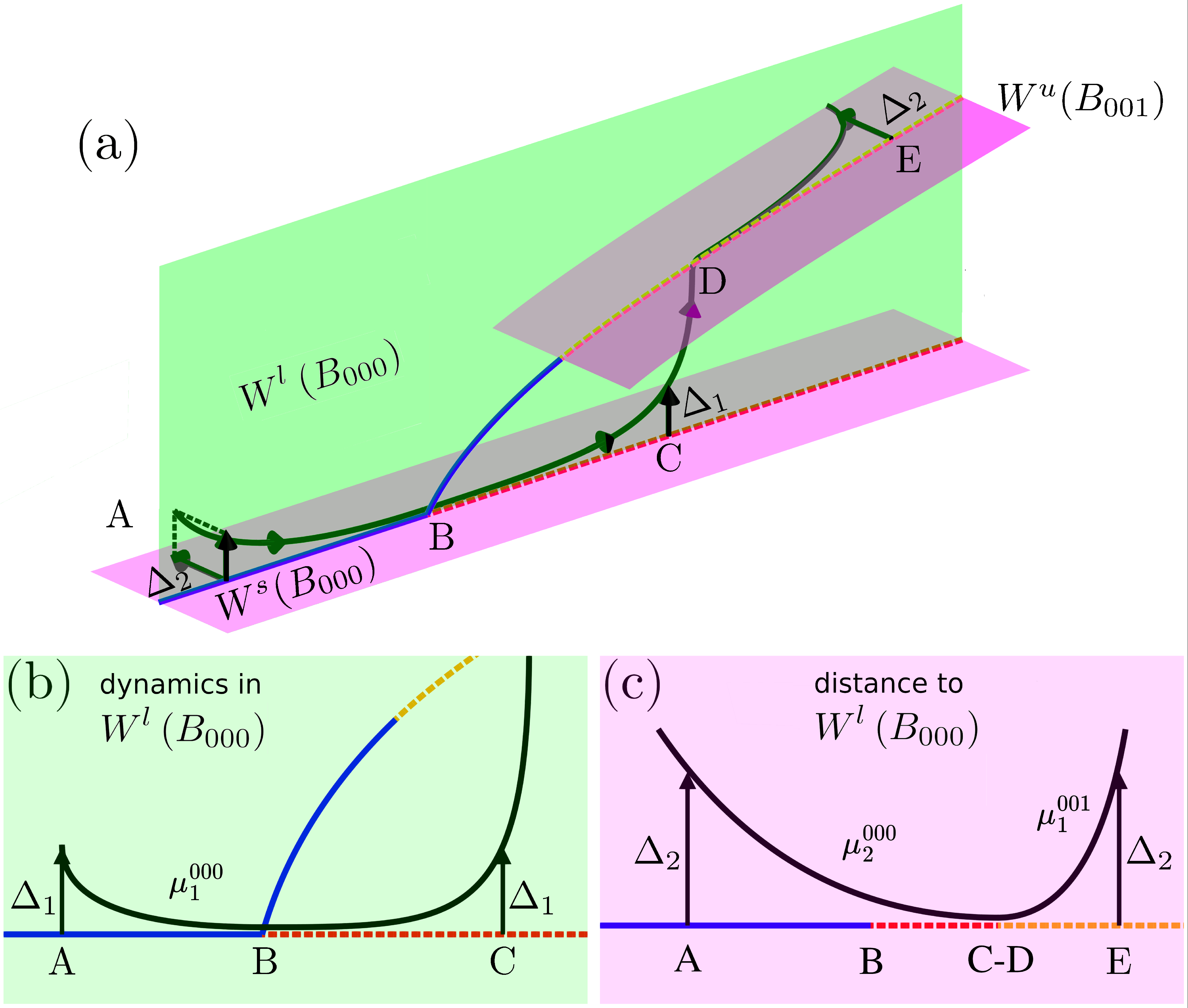}
    \caption{Schematic description of CC jumps for the branches $B_{000}$ and $B_{001}$ for the case of three lasers; see details in the text. 
    }
    \label{fig:3D}
\end{figure}

(q4) Figure~\ref{fig:3D} illustrates the geometric mechanism of CC as the interplay between the branches $B_I$, their stable and unstable manifolds, and the heteroclinic connections between them. 
The figure shows the ``leading manifold" $W^l(B_{000})$ of the branch $B_{000}$, which is defined by the most unstable (least stable) direction of $B_{000}$ \footnote{more precisely, a perturbation of the family of leading stable, centre and leading unstable manifolds of the corresponding equilibria of the layer system parameterized by $\omega$}.
The manifold $W^l(B_{000})$ contains $B_{001}$. 
Since the CC solution is repelled from $B_{000}$ along $W^l(B_{000})$, it is attracted to the branch $B_{001}$, even though $B_{001}$ is unstable. 
Further, the leading unstable manifold $W^u(B_{001})$ of $B_{001}$ is shown in magenta in Fig.~\ref{fig:3D}. The time the system spends near $B_{001}$ depends on the initial distance of the system to $W^l(B_{000})$ and the rate at which the instability develops along $B_{001}$. 

(q5) Now we explain what determines the duration of the slow motions along the unstable branches, see Fig.~\ref{fig:3D}. The part AB of the trajectory shows the approaching to $B_{000}$ and a drift along the stable part of this branch. The duration and rate of convergence to $B_{000}$ determines how long the trajectory remains close to $B_{000}$ during the BC phase. The ABC phase is a standard delayed exchange stability 
\cite{Baer1989,suPhenomenonDelayedBifurcation2001} with the additional simplification that the state $B_{000}$ does not depend on $\omega$. Thus, the jumping condition at $\omega = \omega_{C}$ is given by
 \begin{align}
     \int_{\omega_A}^{\omega_{C}} 
     \frac{\mu_{1}^{000}(\omega)}{\omega + f(X_{000})} \, \text{d} \omega = 0, 
    \label{eq:jump1}
\end{align}
where $\mu_{1}^{000}$ is the leading unstable eigenvalue at 
$B_{000}$, $X_{000}=0$ is the corresponding mean-field.
The theoretically predicted  jump point at $\omega = \omega_C$ for the case $N=3$ (orange dot in Fig.~\ref{fig:cm_comparison}(b)) agrees 
with the actual jump 
of the black trajectory.

We will now explain the mechanism behind the other jump points using the slow motion at the branch $B_{001}$, see Fig.~\ref{fig:3D}. 
The distance $\Delta_2(t)$ from the leading manifold $W^l(B_{000})$ (green in the figure) decreases during the phase ABC along the direction that is transverse to $W^l(B_{000})$ (determined by the second largest eigenvalue $\mu_2^{000}$), and the repulsion during the phase DE along the most unstable direction of $B_{001}$ (determined by the largest eigenvalue $\mu_1^{001}$). 
As a result, the conditions for the second jump point is
\begin{align}
    \int_{\omega_A}^{\omega_{C}} \frac{\mu_2^{000}(\omega)}{\omega + f(X_{000})} \, \text{d} \omega + \int_{\omega_{C}}^{\omega_{E}} \frac{\mu_1^{001}(\omega)}{\omega + f(X_{001})} \, \text{d} \omega = 0. 
    \label{eq:jump2}
\end{align}
Here we used $\omega_C\approx\omega_D$ as the jump occurs on the fast timescale. 
Supplemental material \cite{supplem} provides more detail.

In summary, we have presented the explanation for the emergence and the main ingredients of canard cascading (CC). These ingredients are (i) multiple branches of low-dimensional critical manifolds. The number of such branches grows extensively with the size of the network. (ii)  Robust heteroclinic connections, which allow for fast transitions between unstable branches. (iii) Delayed exchange of stability on the critical manifold allows the appearance of canard solutions following the unstable parts of the branches.
(iv) Finally, the heterogeneity of the individual elements is necessary for the 
switching behaviour.  

While our study focuses on coupled semiconductor lasers and is directly related to an experimentally observed effect, CC is expected in a variety of other setups, such as neural or ecological systems. From a modelling point of view, the important elements seem to be very general: global coupling to an active low-pass filter \cite{ciszakCollectiveCanardExplosions2021} and heterogeneity of interacting elements. 

\begin{acknowledgments}
SY, RB, and JK acknowledge funding by the "Deutsche Forschungsgemeinschaft” (DFG), Project 411803875.
\end{acknowledgments}


%

\onecolumngrid

\section{A.I. Stability analysis of the critical manifold}

In the main part of the paper, we have introduced the critical manifold (see Eqs. (3) and (4) in the manuscript), which consists of branches $B_I$. In this section, we derive the stability conditions for $B_I$. For this, we linearize the fast system along these branches and study the linear stability of the linearizations obtained. 

Let us introduce the following notations  for the right-hand sides of the fast system
\begin{align}
    h_{i, 1}(x_i, y_i) &\coloneqq x_i (y_i - 1), \nonumber\\
    h_{i, 2}(x_i, y_i, \omega) &\coloneqq \gamma [\delta_i - y_i + k( \omega + f(X)) - x_i y_i], \nonumber 
\end{align}
where $ X = \frac{1}{N}  \sum_i x_i$ and $f(X) = A \ln\left(1 + \frac{\alpha}{N}  \sum_i x_i\right)$.
Then the Jacobian of the fast system is
\begin{equation}
\label{general Jacobian of fast-variables}
\text{D}_{\mathbf{x}, \mathbf{y}} \mathbf{h} = \begin{bmatrix}
\frac{\partial h_{1, 1}}{\partial x_1} & \frac{\partial h_{1, 1}}{\partial y_1} & \cdots & \frac{\partial h_{1, 1}}{\partial x_N} & \frac{\partial h_{1, 1}}{\partial y_N} \\
\frac{\partial h_{1, 2}}{\partial x_1} & \frac{\partial h_{1, 2}}{\partial y_1} & \cdots & \frac{\partial h_{1, 2}}{\partial x_N} & \frac{\partial h_{1, 2}}{\partial y_N} \\
\vdots & \vdots & \ddots & \vdots & \vdots \\
\frac{\partial h_{N, 1}}{\partial x_1} & \frac{\partial h_{N, 1}}{\partial y_1} & \cdots & \frac{\partial h_{N, 1}}{\partial x_N} & \frac{\partial h_{N, 1}}{\partial y_N} \\
\frac{\partial h_{N, 2}}{\partial x_1} & \frac{\partial h_{N, 2}}{\partial y_1} & \cdots & \frac{\partial h_{N, 2}}{\partial x_N} & \frac{\partial h_{N, 2}}{\partial y_N} 
\end{bmatrix}.
\end{equation}
The Jacobian contains 
 the following $2 \times 2$-blocks 
\begin{equation}
\label{eq: blocks on diagonal}
\begin{pmatrix}
\frac{\partial h_{i, 1}}{\partial x_i} & \frac{\partial h_{i, 1}}{\partial y_i} \\
\frac{\partial h_{i, 2}}{\partial x_i} & \frac{\partial h_{i, 2}}{\partial y_i}
\end{pmatrix} 
= 
\begin{pmatrix}
y_i - 1 & x_i \\
\gamma \left[ \frac{k A \alpha}{N (1 + \alpha X)} - y_i \right] & - \gamma (x_i + 1)
\end{pmatrix}
\end{equation}
along the diagonal, and all the remaining $2 \times 2$ blocks have the form
\begin{equation}
\label{eq: remaining blocks}
\begin{pmatrix}
\frac{\partial h_{i, 1}}{\partial x_j} & \frac{\partial h_{i, 1}}{\partial y_j} \\
\frac{\partial h_{i, 2}}{\partial x_j} & \frac{\partial h_{i, 2}}{\partial y_j}
\end{pmatrix} =  \begin{pmatrix}
0 & 0 \\
\gamma \frac{k A \alpha}{N (1 + \alpha X)} & 0 
\end{pmatrix},\quad i\ne j.
\end{equation}
Now consider the branch $B_I = B_{a_1\dots a_N}$, where $a_1=1$ if the laser $i$ is on and $a_i=0$ if the laser $i$ is off. Then $N_+ = \sum_i a_i$ is the number of lasers switched on for the solution on this branch.
The corresponding mean field $X_I $ satisfies 
\begin{align*}
    X_I = \frac{1}{N}  \sum_i x_i =
    \frac{1}{N}  \sum_{i\in I_+}  \left( \delta_i - 1 + k [\omega + f(X_I)] \right)
    = 
    \frac{N_{+}}{N} \left[ \bar{\delta}_I - 1 + k( \omega + f(X_I)) \right],
\end{align*}
which leads to the equation (4) from the main part of the manuscript. 

Restricting the Jacobian to the branch $B_I$, we obtain 
\begin{align*}
 \mathbf{J}_{\text{c}} :=
\begin{pmatrix}
0 & 0 \\
\gamma \frac{k A \alpha}{N (1 + \alpha X_{I})} & 0 
\end{pmatrix}
\end{align*}
for the non-diagonal blocks (\ref{eq: remaining blocks}). For the diagonal blocks  (\ref{eq: blocks on diagonal}), we obtain the following two cases depending on whether the laser $i$ is on or off: 
\begin{align}
\mathbf{J}_{\text{on}, i} :=
\begin{pmatrix}
0 & \delta_i - 1 + k[\omega +f(X_I)] \\
\gamma \left[ \frac{k A \alpha}{N (1 + \alpha X_{I})} - 1 \right] & - \gamma \left(\delta_i + k[\omega +f(X_I)] \right)
\end{pmatrix},
\quad \text{if} \quad  a_i =1
 \nonumber
\end{align}
and
\begin{align}
\mathbf{J}_{\text{off}, i} :=
\begin{pmatrix}
\delta_i + k[\omega +f(X_I)] - 1 & 0 \\
\gamma \left[ \frac{k A \alpha}{N (1 + \alpha X_{I})} - \delta_i - k[\omega +f(X_I)] \right] & - \gamma 
\end{pmatrix},
\quad \text{if} \quad  a_i =0.
\nonumber
\end{align}

Without loss of generality, let us assume that the first $N_{+}$ lasers on the branch $B_{I}$ are on. Then the Jacobian for $B_{I}$ has the following block structure:
\begin{equation}
\label{eq: Jacobian along any branch}
J_I := \left. {\text{D}_{\mathbf{x}, \mathbf{y}} \mathbf{h}}\right|_{B_{I}} 
= \left[ \begin{array}{c | c}
\mathbf{ON}_{N_{+}} & \mathbf{C}_{N_{+} \times N_{-}} \\
\hline
\mathbf{C}_{N_{-} \times N_{+}} & \mathbf{OFF}_{N_{-}}
\end{array} \right] \in \mathbb{R}^{2N \times 2N},
\end{equation}
where 
\begin{align}
\mathbf{C}_{p \times q} &= 
\begin{bmatrix}
\mathbf{J}_{\text{c}} & \cdots & \mathbf{J}_{\text{c}} \\
\vdots & \ddots & \vdots \\
\mathbf{J}_{\text{c}} & \cdots & \mathbf{J}_{\text{c}}  
\end{bmatrix} 
\in \mathbb{R}^{2p \times 2q}, \\
\mathbf{ON}_{N_{+}} &= 
\begin{bmatrix}
\mathbf{J}_{\text{on}, 1} & \mathbf{J}_{\text{c}} & \cdots & \mathbf{J}_{\text{c}} \\
\mathbf{J}_{\text{c}} & \mathbf{J}_{\text{on}, 2} & \ddots & \vdots \\
\vdots & \ddots & \ddots & \mathbf{J}_{\text{c}}\\
\mathbf{J}_{\text{c}} & \cdots & \mathbf{J}_{\text{c}} & \mathbf{J}_{\text{on}, N_{+}}
\end{bmatrix} 
\in \mathbb{R}^{2N_{+} \times 2N_{+}}, \\
\mathbf{OFF}_{N_{-}} &= 
\begin{bmatrix}
\mathbf{J}_{\text{off}, N_+ + 1} & \mathbf{J}_{\text{c}} & \cdots & \mathbf{J}_{\text{c}} \\
\mathbf{J}_{\text{c}} & \mathbf{J}_{\text{off}, N_+ + 2} & \ddots & \vdots \\
\vdots & \ddots & \ddots & \mathbf{J}_{\text{c}}\\
\mathbf{J}_{\text{c}} & \cdots & \mathbf{J}_{\text{c}} & \mathbf{J}_{\text{off}, N}
\end{bmatrix} \in \mathbb{R}^{2N_{-} \times 2N_{-}}.
\end{align}

\subsection{A.I.1. Symmetric case}

Here we consider the symmetric case, i.e. all lasers are identical with $\delta = \delta_i$ for all $i$. This case could equivalently be called "homogeneous", but we use "symmetric" throughout for consistency.
Then the corresponding blocks $\mathbf{J}_{\text{on}, i} = \mathbf{J}_{\text{on}}$ and 
$\mathbf{J}_{\text{off}, i} = \mathbf{J}_{\text{off}}$ become identical, where
\begin{align}
\mathbf{J}_{\text{on}} = 
\begin{pmatrix}
0 & \delta - 1 + k[\omega +f(X_I)] \\
\gamma \left[ \frac{k A \alpha}{N (1 + \alpha X_{I})} - 1 \right] & - \gamma \left(\delta + k[\omega +f(X_I)]\right)
\end{pmatrix},
\end{align}
\begin{align}
\mathbf{J}_{\text{off}} = 
\begin{pmatrix}
\delta + k[\omega +f(X_I)] - 1 & 0 \\
\gamma \left[ \frac{k A \alpha}{N (1 + \alpha X_{I})} - \delta - k[\omega +f(X_I)] \right] & - \gamma 
\end{pmatrix}
.
\end{align}
This enables us to find the eigenvalues analytically. 
To formulate our results in a more structured way, we formulate the next result as a lemma. 

\textbf{Lemma 1 [Dimension reduction of the eigenvalue problem].}
{\it 
\label{lemma: reduction lemma}
The eigenvalue problem 
\begin{equation}
\label{eq: original eigenvalue problem}
\mathbf{J}_{I} \mathbf{v} = \lambda \mathbf{v}
\end{equation}
can be reduced in the symmetric case and for $1 \leq N_{+} < N $ to the following three eigenvalue problems
\begin{equation}
\label{eq: general reduced eigenvalue problem ON}
[ \mathbf{J}_{\textnormal{on}} - \mathbf{J}_{\textnormal{c}} ] \mathbf{z}_{\textnormal{on}} = \lambda_{\textnormal{on}} \mathbf{z}_{\textnormal{on}}, 
\end{equation}
\begin{equation}
\label{eq: general reduced eigenvalue problem OFF}
[ \mathbf{J}_{\textnormal{off}} - \mathbf{J}_{\textnormal{c}} ] \mathbf{z}_{\textnormal{off}} = \lambda_{\textnormal{off}} \mathbf{z}_{\textnormal{off}},
\end{equation}
and 
\begin{equation}
\label{eq: reduced eigenvaue problem "remaining case"}
\begin{bmatrix}
[\mathbf{J}_{\textnormal{on}} + (N_{+} - 1) \mathbf{J}_{\textnormal{c}}] & N_{-} \mathbf{J}_{\textnormal{c}} \\
N_{+} \mathbf{J}_{\textnormal{c}} & [\mathbf{J}_{\textnormal{off}} + (N_{-} - 1) \mathbf{J}_{\textnormal{c}}]
\end{bmatrix} \begin{pmatrix}
\mathbf{a} \\
\mathbf{b}
\end{pmatrix} = \lambda \begin{pmatrix}
\mathbf{a} \\
\mathbf{b}
\end{pmatrix}.
\end{equation}
\noindent More specifically, the following are the eigenvector-eigenvalue pairs of (\ref{eq: original eigenvalue problem})
\begin{align*}
\left(  ( \mathbf{v}_{k} \otimes \mathbf{z}_{\textnormal{on}} ),\ \lambda_{\textnormal{on}}\right), \quad k=1, \cdots, N_{+} - 1,\\
\left( ( \mathbf{w}_k \otimes \mathbf{z}_{\textnormal{off}} ),\  \lambda_{\textnormal{off}} \right),\quad k = 1, \cdots, N_{-} - 1
\end{align*}
with 
\begin{align*}
    \mathbf{v}_{k} = (\omega_{k}, \omega_{k}^{2}, \cdots, \omega_{k}^{N_{+}}, 0, \cdots, 0)^T , \\
    \mathbf{w}_{k} = (0, \cdots, 0, \mu_{k}, \mu_{k}^{2}, \cdots, \mu_{k}^{N_{-}})^T 
\end{align*}
and
\begin{align*}
    \omega_{k}^{j} = \exp\left(\frac{2\pi i}{N_{+}} jk\right),\quad  \mu_{k}^{j} = \exp\left(\frac{2\pi i}{N_{-}} jk\right), 
\end{align*}
 if  $\left( \mathbf{z}_{\textnormal{on}}, \lambda_{\textnormal{on}}\right)$ or $\left( \mathbf{z}_{\textnormal{off}}, \lambda_{\textnormal{off}}\right)$  are  the eigenvector-eigenvalue pairs of (\ref{eq: general reduced eigenvalue problem ON}) or (\ref{eq: general reduced eigenvalue problem OFF}) .

Likewise, 
\begin{align*}
    \left( 
    \begin{pmatrix}
        \mathbb{1}_{N_{+}} \otimes \mathbf{a} \\ 
        \mathbb{1}_{N_{-}} \otimes \mathbf{b}
    \end{pmatrix}
    ,
    \lambda 
    \right) 
\end{align*}
is the eigenvector-eigenvalue pair of (\ref{eq: original eigenvalue problem}) if $ ((\mathbf{a}, \mathbf{b}), \lambda)$  is the eigenvector-eigenvalue pair of (\ref{eq: reduced eigenvaue problem "remaining case"}). In this way, we reduce the eigenvalues  problem \eqref{eq: original eigenvalue problem} to the low-dimensional problems  \eqref{eq: general reduced eigenvalue problem ON}, \eqref{eq: general reduced eigenvalue problem OFF}, and \eqref{eq: reduced eigenvaue problem "remaining case"}. 
}

{\it Proof.} 
We will prove the reduction to (\ref{eq: general reduced eigenvalue problem ON}) with the family of vectors $\mathbf{v}_{k}$. 
Consider $\mathbf{z_\textnormal{on}} \ne {\mathbf{0}}$ to be the eigenvector of the reduced problem \eqref{eq: reduced eigenvalue problem ON, N+ = N} with the eigenvalue $\lambda_\textnormal{on}$. 
The identity $\mathbf{J}_{I} ( \mathbf{v}_{k} \otimes \mathbf{z_\textnormal{on}} ) = \lambda_\textnormal{on} ( \mathbf{v}_{k} \otimes \mathbf{z_\textnormal{on}} )$ follows 
from the following calculations:
\allowdisplaybreaks
\begin{align}
\mathbf{J}_{I} ( \mathbf{v}_{k} \otimes \mathbf{z}_\textnormal{on} ) 
&= \left[ \begin{array}{c | c}
\mathbf{ON}_{N_{+}} & \mathbf{C}_{N_{+} \times N_{-}} \nonumber \\
\hline
\mathbf{C}_{N_{-} \times N_{+}} & \mathbf{OFF}_{N_{-}}
\end{array} \right] (\omega_{k}, \omega_{k}^{2}, \cdots, \omega_{k}^{N_{+}}, 0, \cdots, 0)^T \otimes \mathbf{z}_\textnormal{on} \nonumber \\ 
&= \left[ \begin{array}{c | c}
\begin{array}{c c c c}
\mathbf{J}_{\text{on}} & \mathbf{J}_{\text{c}} & \cdots & \mathbf{J}_{\text{c}} \\
\mathbf{J}_{\text{c}} & \mathbf{J}_{\text{on}} & \ddots & \vdots \\
\vdots & \ddots & \ddots & \mathbf{J}_{\text{c}}\\
\mathbf{J}_{\text{c}} & \cdots & \mathbf{J}_{\text{c}} & \mathbf{J}_{\text{on}}
\end{array} & \mathbf{C}_{N_{+} \times N_{-}} \\
\hline
\mathbf{C}_{N_{-} \times N_{+}} & \mathbf{OFF}_{N_{-}}
\end{array} \right] \begin{pmatrix}
\omega_{k} \\
\omega_{k}^{2}\\
\vdots \\
\omega_{k}^{N_{+}} \\
0 \\
\vdots \\
0
\end{pmatrix} \otimes \mathbf{z}_\textnormal{on} \nonumber \\
&= \begin{pmatrix}
&\mathbf{J}_{\text{on}} \omega_{k} &+ &\sum\limits_{j \neq 1} \omega_{k}^{j} \mathbf{J}_{\text{c}} \\
&\mathbf{J}_{\text{on}} \omega_{k}^{2} &+ &\sum\limits_{j \neq 2} \omega_{k}^{j} \mathbf{J}_{\text{c}} \\
& &\vdots & \\
&\mathbf{J}_{\text{on}} \omega_{k}^{N_{+}} &+ &\sum\limits_{j \neq N_{+}} \omega_{k}^{j} \mathbf{J}_{\text{c}} \\
& &\sum\limits_{j=1}^{N_{+}} \omega_{k}^{j} \mathbf{J}_{\text{c}} & \\
& &\vdots &\\
& &\sum\limits_{j=1}^{N_{+}} \omega_{k}^{j} \mathbf{J}_{\text{c}} &
\end{pmatrix} \otimes \mathbf{z}_\textnormal{on} 
= \begin{pmatrix}
&\mathbf{J}_{\text{on}} \omega_{k} &- & \mathbf{J}_{\text{c}} \omega_{k} &+ &\mathbf{J}_{\text{c}} \sum\limits_{j=1}^{N_{+}} \omega_{k}^{j} \\
&\mathbf{J}_{\text{on}} \omega_{k}^{2} &- &\mathbf{J}_{\text{c}} \omega_{k}^{2} &+ & \mathbf{J}_{\text{c}} \sum\limits_{j=1}^{N_{+}} \omega_{k}^{j}  \\
& & &\vdots & &  \\
&\mathbf{J}_{\text{on}} \omega_{k}^{N_{+}} &- &\mathbf{J}_{\text{c}} \omega_{k}^{N_{+}} &+ & \mathbf{J}_{\text{c}} \sum\limits_{j=1}^{N_{+}} \omega_{k}^{j} \\
& & &0 & & \\
& & &\vdots & & \\
& & &0 & & 
\end{pmatrix} 
\otimes \mathbf{z}_\textnormal{on} \nonumber \\
&= \begin{pmatrix}
\omega_{k} [\mathbf{J}_{\text{on}} - \mathbf{J}_{\text{c}} ] \mathbf{z}_\textnormal{on} \\
\omega_{k}^{2} [\mathbf{J}_{\text{on}} - \mathbf{J}_{\text{c}} ] \mathbf{z}_\textnormal{on} \\
\vdots \\
\omega_{k}^{N_{+}} [\mathbf{J}_{\text{on}} - \mathbf{J}_{\text{c}} ] \mathbf{z}_\textnormal{on} \\
0 \\
\vdots \\
0 
\end{pmatrix} 
= 
\mathbf{v}_{k} \otimes [ \mathbf{J}_{\text{on}} - \mathbf{J}_{\text{c}} ] \mathbf{z}_\textnormal{on}
= 
\mathbf{v}_{k} \otimes (\lambda_\textnormal{on} \mathbf{z}_\textnormal{on})
=
\lambda_\textnormal{on}  (\mathbf{v}_{k} \otimes \mathbf{z}_\textnormal{on})
. \label{left-hand side k != 0}
\end{align}
The proof of $\mathbf{J}_{I} ( \mathbf{w}_{k} \otimes \mathbf{z_\textnormal{off}} ) = \lambda_\textnormal{off} ( \mathbf{w}_{k} \otimes \mathbf{z_\textnormal{off}} )$ can be done analogously.

For the remaining case (\ref{eq: reduced eigenvaue problem "remaining case"}), let us consider
\begin{align*}
\mathbf{v} = (\underbrace{\mathbf{a}, \cdots, \mathbf{a}}_{N_{+}}, 
                \underbrace{\mathbf{b}, \cdots, \mathbf{b}}_{N_-})^T = 
    \begin{pmatrix}
        \mathbb{1}_{N_{+}} \otimes \mathbf{a} \\ 
        \mathbb{1}_{N_{-}} \otimes \mathbf{b}
    \end{pmatrix}.
\end{align*}
Direct calculations lead to 
\begin{align*}
\mathbf{J}_{I} \mathbf{v} &= \left[ \begin{array}{c | c}
\mathbf{ON}_{N_{+}} & \mathbf{C}_{N_{+} \times N_{-}} \\
\hline
\mathbf{C}_{N_{-} \times N_{+}} & \mathbf{OFF}_{N_{-}}
\end{array} \right] 
\begin{pmatrix}
        \mathbb{1}_{N_{+}} \otimes \mathbf{a} \\ 
        \mathbb{1}_{N_{-}} \otimes \mathbf{b}
    \end{pmatrix}
\\
&= \left[ \begin{array}{c | c}
\begin{array}{cccc}
\mathbf{J}_{\text{on}} & \mathbf{J}_{\text{c}} & \cdots & \mathbf{J}_{\text{c}} \\
\mathbf{J}_{\text{c}} & \mathbf{J}_{\text{on}} & \ddots & \vdots \\
\vdots & \ddots & \ddots & \mathbf{J}_{\text{c}}\\
\mathbf{J}_{\text{c}} & \cdots & \mathbf{J}_{\text{c}} & \mathbf{J}_{\text{on}}
\end{array} &
\begin{array}{cccc}
\mathbf{J}_{\text{c}} & \cdots & \cdots & \mathbf{J}_{\text{c}} \\
\vdots & \ddots & \ddots & \vdots \\
\vdots & \ddots & \ddots & \vdots \\
\mathbf{J}_{\text{c}} & \cdots & \cdots & \mathbf{J}_{\text{c}}
\end{array} \\
\hline
\begin{array}{cccc}
\mathbf{J}_{\text{c}} & \cdots & \cdots & \mathbf{J}_{\text{c}} \\
\vdots & \ddots & \ddots & \vdots \\
\vdots & \ddots & \ddots & \vdots \\
\mathbf{J}_{\text{c}} & \cdots & \cdots & \mathbf{J}_{\text{c}}
\end{array} & \begin{array}{cccc}
\mathbf{J}_{\text{off}} & \mathbf{J}_{\text{c}} & \cdots & \mathbf{J}_{\text{c}} \\
\mathbf{J}_{\text{c}} & \mathbf{J}_{\text{off}} & \ddots & \vdots \\
\vdots & \ddots & \ddots & \mathbf{J}_{\text{c}}\\
\mathbf{J}_{\text{c}} & \cdots & \mathbf{J}_{\text{c}} & \mathbf{J}_{\text{off}}
\end{array}
\end{array} \right] \begin{pmatrix}
\mathbf{a} \\
\vdots \\
\mathbf{a} \\
\mathbf{b} \\
\vdots \\
\mathbf{b}
\end{pmatrix} \\
\\
&= \begin{pmatrix}
&[\mathbf{J}_{\text{on}} + (N_{+} - 1) \mathbf{J}_{\text{c}}] \mathbf{a} &+ &N_{-} \mathbf{J}_{\text{c}} \mathbf{b} \\
& &\vdots & \\
&[\mathbf{J}_{\text{on}} + (N_{+} - 1) \mathbf{J}_{\text{c}}] \mathbf{a} &+ &N_{-} \mathbf{J}_{\text{c}} \mathbf{b} \\
&N_{+} \mathbf{J}_{\text{c}} \mathbf{a} &+ &[\mathbf{J}_{\text{off}} + (N_{-} - 1) \mathbf{J}_{\text{c}}] \mathbf{b} \\
& &\vdots &\\
&N_{+} \mathbf{J}_{\text{c}} \mathbf{a} &+ &[\mathbf{J}_{\text{off}} + (N_{-} - 1) \mathbf{J}_{\text{c}}] \mathbf{b} \\
\end{pmatrix} 
\overset{\eqref{eq: reduced eigenvaue problem "remaining case"}}{=}
\begin{pmatrix}
\lambda  \mathbf{a} \\
\vdots  \\
\lambda  \mathbf{a} \\
\lambda  \mathbf{b}  \\
\vdots \\
\lambda  \mathbf{b}  \\
\end{pmatrix} 
=
\lambda \mathbf{v}.
\end{align*}\\
\textit{Proof is complete.}

Note that Lemma 1 does not deal with the cases $N_{+}=0$ and $N_{+}=N$. The following lemmas treat these cases and they are given without proof, since their proof is analogous to Lemma 1. 

\textbf{Lemma 2 [Dimension reduction of the eigenvalue problem (case $N_+=0$)].}
\textit{The eigenvalue problem  \eqref{eq: original eigenvalue problem} in the case $N_{+} = 0$
can be reduced to the following two eigenvalue problems
\begin{equation}
\label{eq: reduced eigenvalue problem OFF, N+ = 0}
[ \mathbf{J}_{\textnormal{off}} - \mathbf{J}_{\textnormal{c}} ] \mathbf{z} = \lambda \mathbf{z}, 
\end{equation}
\begin{equation}
    \label{eq: off part of reduced remaining case}
    [\mathbf{J}_{\textnormal{off}} + (N - 1) \mathbf{J}_{\textnormal{c}}] \mathbf{b} = \lambda \mathbf{b}.
\end{equation}
\noindent More specifically, the following are the eigenvector-eigenvalue pairs of (\ref{eq: original eigenvalue problem})
\begin{align*}
\left( ( \mathbf{w}_k \otimes \mathbf{z}_{\textnormal{off}} ),\  \lambda_{\textnormal{off}} \right),\quad k = 1, \cdots, N - 1
\end{align*}
with 
\begin{align*}
    \mathbf{w}_{k} = (\mu_{k}, \mu_{k}^{2}, \cdots, \mu_{k}^{N})^T 
\end{align*}
and
\begin{align*}
    \mu_{k}^{j} = \exp\left(\frac{2\pi i}{N} jk\right), 
\end{align*}
if $\left( \mathbf{z}_{\textnormal{off}}, \lambda_{\textnormal{off}}\right)$ are the eigenvector-eigenvalue pairs of (\ref{eq: reduced eigenvalue problem OFF, N+ = 0}).
Likewise, 
\begin{align*}
    \left(\mathbb{1}_{N} \otimes \mathbf{b}, \lambda \right) 
\end{align*}
is the eigenvector-eigenvalue pair of (\ref{eq: original eigenvalue problem}) if $(\mathbf{b}, \lambda)$ is the eigenvector-eigenvalue pair of (\ref{eq: off part of reduced remaining case}).}

\textbf{Lemma 3 [Dimension reduction of the eigenvalue problem (case $N_+=N$)].}
\textit{The eigenvalue problem \eqref{eq: original eigenvalue problem} in the case $N_{+} = N$ can be reduced to the following two eigenvalue problems
\begin{equation}
\label{eq: reduced eigenvalue problem ON, N+ = N}
[ \mathbf{J}_{\textnormal{on}} - \mathbf{J}_{\textnormal{c}} ] \mathbf{z} = \lambda \mathbf{z}, 
\end{equation}
\begin{equation}
    \label{eq: on part of reduced remaining case}
    [\mathbf{J}_{\textnormal{on}} + (N - 1) \mathbf{J}_{\textnormal{c}}] \mathbf{a} = \lambda \mathbf{a}.
\end{equation}
\noindent More specifically, the following are the eigenvector-eigenvalue pairs of (\ref{eq: original eigenvalue problem})
\begin{align*}
\left(  ( \mathbf{v}_{k} \otimes \mathbf{z}_{\textnormal{on}} ),\ \lambda_{\textnormal{on}}\right), \quad k=1, \cdots, N - 1
\end{align*}
with 
\begin{align*}
    \mathbf{v}_{k} = (\omega_{k}, \omega_{k}^{2}, \cdots, \omega_{k}^{N})^T    
\end{align*}
and
\begin{align*}
    \omega_{k}^{j} = \exp\left(\frac{2\pi i}{N} jk\right), 
\end{align*}
 if $\left( \mathbf{z}_{\textnormal{on}}, \lambda_{\textnormal{on}}\right)$ are the eigenvector-eigenvalue pairs of (\ref{eq: reduced eigenvalue problem ON, N+ = N}).
Likewise, 
\begin{align*}
    \left( \mathbb{1}_{N} \otimes \mathbf{a}, \lambda \right) 
\end{align*}
is the eigenvector-eigenvalue pair of (\ref{eq: original eigenvalue problem}) if $ (\mathbf{a}, \lambda)$ is the eigenvector-eigenvalue pair of (\ref{eq: on part of reduced remaining case}).}
After reducing the original problem \eqref{eq: original eigenvalue problem} for the stability of the critical manifold into low-dimensional problems in Lemmas 1-3, we present the eigenvalues of these reduced problems in  lemma 4.

\vbox{
\textbf{Lemma 4 [Eigenvalues].} 
\textit{
    \begin{itemize}
        \item[1)] The solutions of the eigenvalue problem 
            \begin{align}
                [ \mathbf{J}_{\textnormal{on}} - \mathbf{J}_{\textnormal{c}} ] \mathbf{z} = \lambda  \mathbf{z} \nonumber
            \end{align}
            are given by the solutions of 
                $$\lambda^2 + \gamma (x_{I} + 1) \lambda + x_{I} \gamma = 0,
                \quad \textnormal{where} \quad 
                x_{I} := \delta -1+k[\omega +f(X_I)].$$
        \item[2)] The solutions of the eigenvalue problem 
            \begin{align}
            [ \mathbf{J}_{\textnormal{off}} - \mathbf{J}_{\textnormal{c}} ] \mathbf{z} = \lambda \mathbf{z} \nonumber
            \end{align}
            are $\lambda_1 = - \gamma$ and $\lambda_2 = x_{I}$.
        \item[3)] The solutions of the eigenvalue problem 
            \begin{align}
            [\mathbf{J}_{\textnormal{on}} + (N - 1) \mathbf{J}_{\textnormal{c}}] \mathbf{a} = \lambda \mathbf{a} \nonumber
            \end{align}
            are given as solutions of 
                $$\lambda^2 + \gamma (x_{I} + 1) \lambda - x_{I} \gamma (\xi_I - 1) = 0, \quad  
                \textnormal{where} \quad \xi_I \coloneqq \frac{k A \alpha}{ 1 + \alpha X_I}.
                $$
        \item[4)] 
            The eigenvalue problem \begin{align}
            [\mathbf{J}_{\textnormal{off}} + (N - 1) \mathbf{J}_{\textnormal{c}}] \mathbf{b} = \lambda \mathbf{b} \nonumber
            \end{align}
            has the same eigenvalues as 2), i.e. $\lambda_1 = - \gamma$ and $\lambda_2 = x_{I}$.
        \item[5)] 
            The eigenvalue problem 
            \begin{align}
            \begin{bmatrix}
            [\mathbf{J}_{\textnormal{on}} + (N_{+} - 1) \mathbf{J}_{\textnormal{c}}] & N_{-} \mathbf{J}_{\textnormal{c}} \\
            N_{+} \mathbf{J}_{\textnormal{c}} & [\mathbf{J}_{\textnormal{off}} + (N_{-} - 1) \mathbf{J}_{\textnormal{c}}]
            \end{bmatrix} \begin{pmatrix}
            \mathbf{a} \\
            \mathbf{b}
            \end{pmatrix} = \lambda \begin{pmatrix}
            \mathbf{a} \\
            \mathbf{b}
            \end{pmatrix} \nonumber
            \end{align}
            has the eigenvalues $\lambda_1 = - \gamma$ and $\lambda_2 = x_{I}$ and, additionally, the solutions of 
            \begin{align}
                \lambda^2 + \gamma (x_I + 1) \lambda - x_I \gamma \left( \frac{N_+}{N} \xi_I - 1 \right) = 0. \nonumber
            \end{align}   
    \end{itemize}
}
}

{\it Proof.} We will only give direct calculations of the characteristic polynomial for cases 3--5, as cases 1 and 2 are even more straightforward.  
\begin{itemize}
    \item[3)]     
        Let us define
        \begin{align}
            \mathbf{A} \coloneqq \mathbf{J}_{\textnormal{on}} + (N - 1) \mathbf{J}_{\textnormal{c}} \,, \nonumber
        \end{align}
        \begin{align}
            \xi_I := \frac{k A \alpha}{1 + \alpha X_I}, \quad \textnormal{and} \quad 
            x_{I} := \delta - 1 + k ( \omega + f(X_I) ). \nonumber
        \end{align}
        Then we obtain the characteristic polynomial $\chi_{\mathbf{A}}(\lambda)$
        as follows:
        \begin{align}
            \chi_{\mathbf{A}}(\lambda) 
            &= \textnormal{det}(\mathbf{A} - \lambda \mathbf{I}) \nonumber \\
            &= \textnormal{det}(\mathbf{J}_{\textnormal{on}} + (N - 1) \mathbf{J}_{\textnormal{c}} - \lambda \mathbf{I}) \nonumber \\
            &= \textnormal{det}\left( 
            \left( \begin{array}{cc}
                0 & x_{I} \\
                \gamma(N^{-1}\xi_I - 1) & -\gamma(x_{I} + 1)
            \end{array} \right) 
            + 
            \left( \begin{array}{cc}
                0 & 0 \\
                (N-1) \gamma N^{-1} \xi_I & 0
            \end{array} \right)
            - 
            \left( \begin{array}{cc}
                \lambda & 0 \\
                0 & \lambda
            \end{array} \right)
            \right) \nonumber \\
            &= \textnormal{det} \left(
            \begin{array}{cc}
                -\lambda & x_{I} \\
                \gamma(\xi_I - 1) & -\gamma(x_{I} + 1) - \lambda
            \end{array}
            \right) \nonumber \\
            &= \lambda^2 + \gamma(x_{I} + 1) \lambda - x_{I} \gamma (\xi_I - 1). \nonumber
        \end{align}
    \item[4)] 
        Let us define 
        \begin{align}
            \mathbf{B} \coloneqq \mathbf{J}_{\textnormal{off}} + (N - 1) \mathbf{J}_{\textnormal{c}} \, , \nonumber
        \end{align}
        and let $\xi_I$, $x_I$ be as in the calculations of $\chi_{\mathbf{A}}(\lambda)$. Then we obtain the characteristic polynomial $\chi_{\mathbf{B}}(\lambda)$
        as follows:
        \begin{align}
            \chi_{\mathbf{B}}(\lambda) 
            &= \textnormal{det}(\mathbf{B} - \lambda \mathbf{I}) \nonumber \\
            &= \textnormal{det}(\mathbf{J}_{\textnormal{off}} + (N - 1) \mathbf{J}_{\textnormal{c}} - \lambda \mathbf{I}) \nonumber \\
            &= \textnormal{det} \left( 
            \left(
            \begin{array}{cc}
                x_I & 0 \\
                \gamma (N^{-1} \xi_I - x_I - 1) & -\gamma
            \end{array}
            \right)
            + 
            \left(
            \begin{array}{cc}
                0 & 0 \\
                (N - 1) \gamma N^{-1} \xi_I & 0
            \end{array}
            \right)
            - 
            \left(
            \begin{array}{cc}
                \lambda & 0 \\
                0 & \lambda
            \end{array}
            \right)
            \right) \nonumber \\
            &= \textnormal{det}
            \left(
            \begin{array}{cc}
                x_I - \lambda & 0 \\
                \gamma(\xi_I - x_I - 1) & -\gamma - \lambda
            \end{array}
            \right) \nonumber \\
            &= (\lambda - x_I) (\lambda - (- \gamma)). \nonumber
        \end{align}
    \item[5)] 
        Again, let $\xi$, $x_I$ be as in the calculations of $\chi_{\mathbf{A}}(\lambda)$. 
        Then the characteristic equation for the case 5) is 
        \begin{align}
            & \textnormal{det}
            \left(
                \begin{array}{cccc}
                    -\lambda & x_I & 0 & 0 \\
                    \gamma (N_+ N^{-1} \xi_I - 1) & -\gamma (x_I + 1) - \lambda & \gamma N_- N^{-1} \xi_I & 0 \\
                    0 & 0 & x_I - \lambda & 0 \\
                    \gamma N_+ N^{-1} \xi_I & 0 & \gamma (N_- N^{-1} \xi_I - x_I - 1) & -\gamma - \lambda   
                \end{array}
            \right) \nonumber \\
            &= (-\gamma - \lambda) \, \textnormal{det}
            \left(
                \begin{array}{ccc}
                    -\lambda & x_I & 0 \\
                    \gamma (N_+ N^{-1} \xi_I - 1) & -\gamma (x_I + 1) - \lambda & \gamma N_- N^{-1} \xi_I \\
                    0 & 0 & x_I - \lambda   
                \end{array}
            \right) \nonumber \\
            &= (-\gamma - \lambda) (x_I - \lambda) \, \textnormal{det} 
            \left(
                \begin{array}{cc}
                    -\lambda & x_I \\
                    \gamma (N_+ N^{-1} \xi_I - 1) & -\gamma (x_I + 1) - \lambda
                \end{array}
            \right) \nonumber \\
            &= (\lambda - (-\gamma)) (\lambda - x_I) 
            \left( 
                \lambda^2 + \gamma (x_I + 1) \lambda - x_I \gamma \left( \frac{N_+}{N} \xi_I - 1 \right)
            \right). \nonumber
        \end{align}
\end{itemize}

\textit{Proof is finished.}

Using the obtained results in Lemma 1 and Lemma 4, we can obtain explicit stability conditions for the critical manifolds.

\textbf{Lemma 5. [Stability of the critical manifolds in symmetric case]} 
\textit{
\label{lemma: stability of critical manifold in symmetric case}
Consider the symmetric case with $\delta_1=\cdots=\delta_N$. Then 
the "off" branch $B_{\mathbf{0}}$ of the critical manifold  is exponentially stable  if and only if 
    \begin{align}
    \label{eq:omega}
        \omega < \frac{1 - \delta}{k}.
    \end{align}
The "all on" branch $B_{\mathbf{1}}$ is exponentially stable if and only if 
    \begin{align*}
        X_{\mathbf{1}} > kA - \frac{1}{\alpha}.
    \end{align*}
All other branches are unstable for all parameter values.
}

{\it Proof.}
We consider different cases:
\begin{itemize}
    \item[1)]$N_{+}=0$: According to Lemmas 2 and 4, the eigenvalues in this case are $\lambda_1 = - \gamma$ and $\lambda_2 = x_{I}$. Since $\gamma > 0$, consequently $\lambda_1 < 0$ always holds true. From $\lambda_2 < 0$, we obtain the first criterion for the exponential stability
    \begin{align}
        x_{I} = \delta + k(\omega + f(X_{I})) - 1 = \delta + k\omega  - 1 < 0, \nonumber 
    \end{align}
    which is equivalent to \eqref{eq:omega}.
    \item[2)]$1 \leq N_{+} < N$: According to Lemmas 1 and 4, the eigenvalues in this case are given by 
    \begin{align}
        \lambda^2 + \gamma (x_I + 1) \lambda + x_I \gamma = 0, \nonumber
    \end{align}
    \begin{align}
        \lambda^2 + \gamma (x_I + 1) \lambda - x_I \gamma \left( \frac{N_+}{N} \xi_I - 1 \right) = 0, 
        \label{eq:Polynom}
    \end{align}
    as well as $\lambda_1 = - \gamma$ and $\lambda_2 = x_{I}$. 
    Since
    \begin{align}
       x_{I} = \frac{N}{N_{+}} X_{+}, 
    \end{align}
    we have $\lambda_2>0$ and these branches are always unstable.
    \item[3)]$N_{+} = N$: In this case the eigenvalues are given by 
    \begin{align}
        \label{polynomial on}
        \lambda^2 + \gamma (x_I + 1) \lambda + x_I \gamma = 0
    \end{align}
    and
    \begin{align}
        \label{polynomial on remaining}
        \lambda^2 + \gamma (x_{I} + 1) \lambda - x_{I} \gamma (\xi_I - 1) = 0. 
    \end{align}
    For the polynomial (\ref{polynomial on}), we see that its coefficients are positive, hence it can have only negative solutions.
    For the polynomial (\ref{polynomial on remaining}), we obtain the following  stability criterion 
    \begin{align}
        - x_I \gamma (\xi_I - 1) > 0, \nonumber 
    \end{align}
    which leads to 
        \begin{align}
        X_{I} > \frac{k A \alpha - 1}{\alpha} = k A - \frac{1}{\alpha} .\nonumber
    \end{align}
\end{itemize}

{\textit{Proof is complete.}}

\section{A.II. Critical manifold in the symmetric case}

\begin{figure}
    \includegraphics[width=\linewidth]{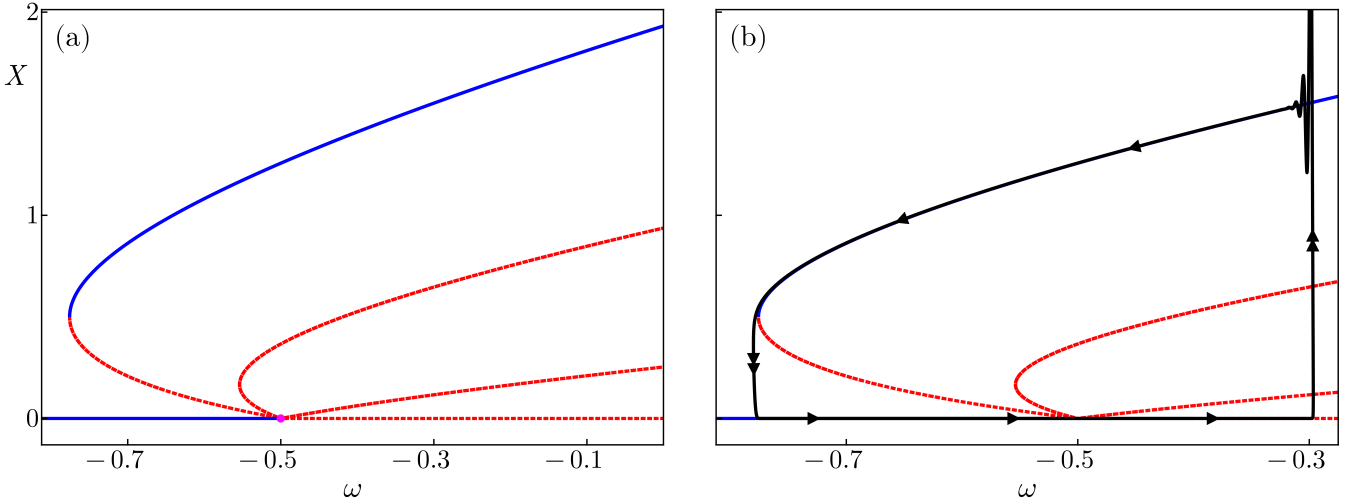}
    \caption{Critical manifolds for the symmetric case. The colors and line styles have the same meaning as in Figs.~3(a,b) of the main manuscript. In addition, the one branching point S is shown in magenta. Parameters: $\delta_1=\delta_2=\delta_3=1.35$, $k=0.7$, $A=k^{-1}$, and $\alpha=2$.
    \label{fig:symmetric_cm_with_dynamics}
    }
\end{figure}

Figure \ref{fig:symmetric_cm_with_dynamics} shows the structure of the critical manifolds for the case of three identical lasers with $\delta_1=\delta_2=\delta_3$. This resembles Figs.~3(a,b) in the main manuscript, which depict the non-symmetric scenario. Stability details for Fig.~\ref{fig:symmetric_cm_with_dynamics} are provided in Lemma 5. The figure demonstrates that the CC phenomenon is absent in the symmetric scenario, while also helping to identify the alterations in the critical manifold's configuration due to the lasers' heterogeneity. In particular, it can be observed that all branches corresponding to the same number of lasers 'on' are projected onto the same line and have the same stability. Furthermore, all branches branch out from a single point S.  

\section{A.III. Heteroclinic connections between branches of the critical manifold \label{sec: heteroclinic connections}}

\subsection{A.III.1 Existence of heteroclinic connections} 

Heteroclinic connections (orbits) for a fixed value of the slow variable $\omega$ are the orbits of the fast subsystem $(\mathbf{x}(t),\mathbf{y}(t))$ connecting its different equilibria. Since these equilibria are on different branches of the critical manifolds $B_I$ and $B_J$, the heteroclinic orbits create connections between the branches. More precisely, we have 
$$
(\mathbf{x}(t),\mathbf{y}(t)) \to B_J \text{ as } t\to +\infty 
\quad \text{ and }  \quad 
(\mathbf{x}(t),\mathbf{y}(t)) \to B_I \text{ as } t\to -\infty .
$$
for a fixed $\omega$ and $\varepsilon=0$ in Eq.~(2) of the main manuscript. 

For small but non-zero $\varepsilon$, the critical manifolds persist as perturbed slow manifolds, and the heteroclinic connections persist between the branches of the slow manifolds (see the robustness discussion in sec.~A.III.2). As a result, a one-parameter family of connecting orbits appears, connecting $B_I$ to $B_J$, parameterised by $\omega$.

The way to represent the heteroclinic connections for a fixed $\omega$ is to treat the equilibria of the fast subsystem as nodes and the heteroclinic connections as directed links of some network. The result is a directed network of heteroclinic connections. Obviously, the equilibria, i.e. the nodes, can be equivalently denoted by the corresponding branches $B_I$ of the critical manifold. The result is a network like the one shown in Fig.~3(c) of the main manuscript.

\begin{figure}[h]
\includegraphics[width=\linewidth]{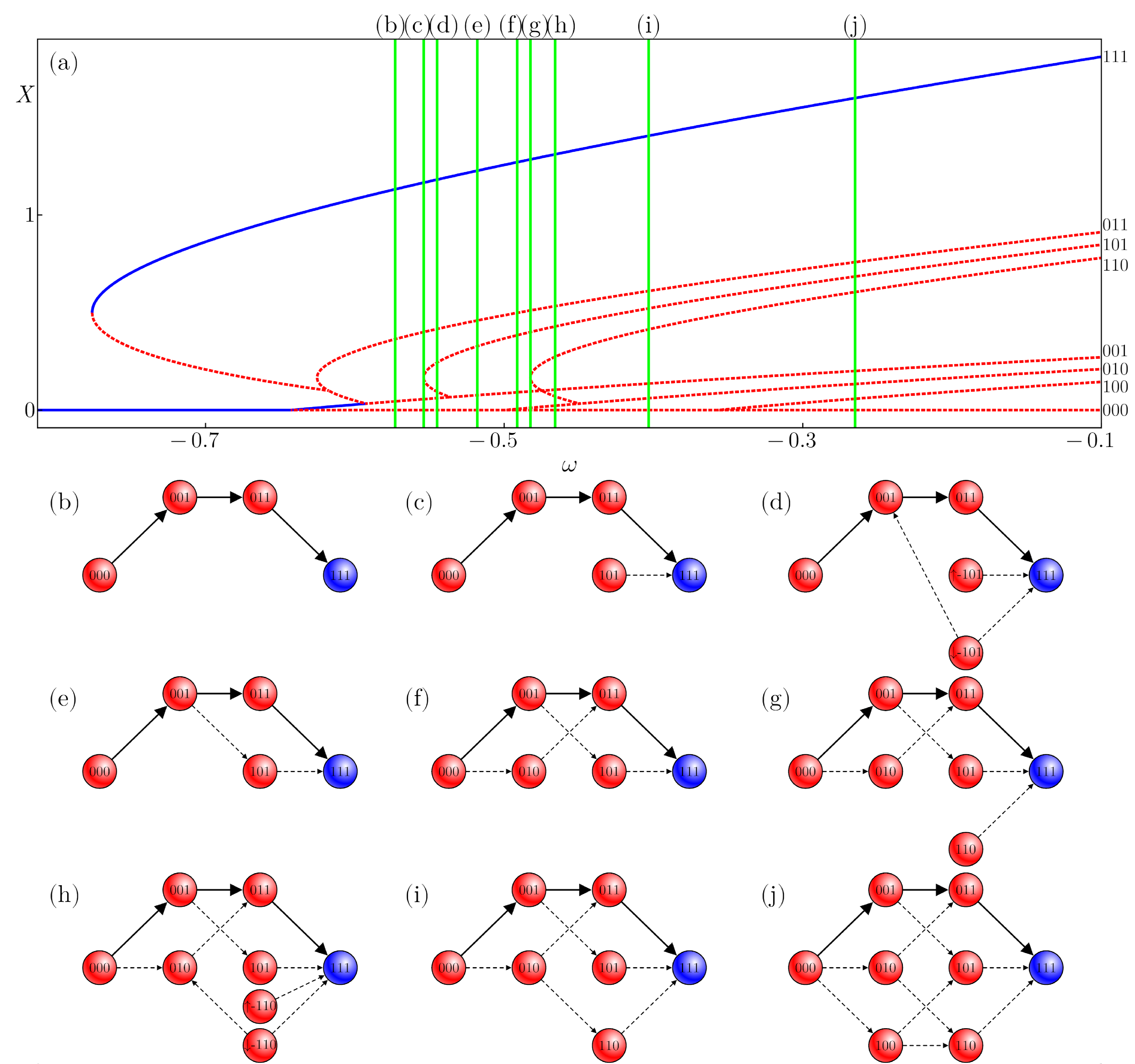}
\caption{Transitions of the heteroclinic connections graph of the non-symmetric case starting at $\omega_b$ where $\omega_b$ is the $\omega$-value of the second branching point along the 001 branch where 001 branch becomes unstable (see (a)). (b) graph before the 101 fold point. (c) graph at the 101 fold point. (d) graph after the 101 fold point before the 101 branching point. (e) graph after the 101 branching point before the second 000 branching point. (f) graph after the second 000 branching point before the 110 fold point. (g) graph at the 110 fold point. (h) graph after the 110 fold point before the 110 branching point. (i) graph after the 110 branching point before the third 000 branching point. (j) graph after the third 000 branching point.
}
\label{fig: heteroclinic network transitions illustration} 
\end{figure}

Figure~\ref{fig: heteroclinic network transitions illustration} shows how the network of heteroclinic connections changes with $\omega$. Here, we do not present a rigorous proof, but rather simple numerical simulations: starting from an unstable equilibrium,  we added a small perturbation of size $10^{-6}$ in the direction of its unstable eigenvectors and obtained the corresponding trajectories from the simulation. The heteroclinic trajectory is recorded when the solution arrives in a $10^{-6}$ neighbourhood of another equilibrium. Note that another argument for the existence of heteroclinic connections is that, due to the general properties of local transcritical bifurcation, there exists a connection close to the branching point \cite{Kuznetsov1995}. As a result, the heteroclinic network evolves with $\omega$ so that it has the form as in Fig.~3(c) of the main manuscript for $\omega > \frac{1 - \delta_1}{k}$.

\subsection{A.III.2 Robustness of heteroclinic connections \label{sec:AIII2}}

In the following we use the following conclusion of the transversality theorem from differential topology
\cite{guilleminDifferentialTopology1974}: If two submanifolds $X$ and $Z$ of $\mathbb{R}^n$ are transversal, then their intersection $X\cap Z$ is a submanifold of $\mathbb{R}^n$ and 
$$
\operatorname{codim}(X \cap Z)=\operatorname{codim} X+\operatorname{codim} Z.
$$
In terms of dimensions, in our case this means
$$
\operatorname{dim}(X \cap Z)= (\operatorname{dim} X + \operatorname{dim} Z) - n.
$$
We will show the following for the heteroclinic connection $p\to q$: 
\begin{align}
    \textnormal{dim}(W^{\textnormal{u}}_{\textnormal{loc}}(p)) + \textnormal{dim}(W^{\textnormal{s}}_{\textnormal{loc}}(q)) > n, \label{genericity_condition}
\end{align}
where $W^{\textnormal{u}}_{\textnormal{loc}}$ is the unstable manifold of equilibrium $p$ and $W^{\textnormal{s}}_{\textnormal{loc}}(q)$ is the stable manifold of equilibrium $q$.
This will imply that if  $W^{\textnormal{u}}_{\textnormal{loc}}(p)$ and $W^{\textnormal{s}}_{\textnormal{loc}}(q)$ intersect transversely, the intersection is a submanifold of dimension $\ge 1$, i.e. there is at least one heteroclinic connection, and this intersection is robust  \cite{guilleminDifferentialTopology1974}. We will not give the proof of the transversality here, which remains a missing ingredient for a rigorous proof of heteroclinic connections. 
Since such a rigorous proof is not our main purpose, we leave it as an open problem.

\textbf{Lemma 6 [Dimensionality of $W^{\textnormal{u}}_{\textnormal{loc}}$].}
{\it
The dimension of the local unstable manifold $W^{\textnormal{u}}_{\textnormal{loc}}$ at the equilibrium of the fast subsystem in the symmetric case is given for $\omega \geq \frac{1 - \delta}{k}$ by
\begin{itemize}
    \item[1)]$N_{+} = 0$:
        \begin{align}
            \textnormal{dim}(W^{\textnormal{u}}_\textnormal{loc}) = N  \nonumber
        \end{align}
    \item[2)]$N_{+}>0$:
        \begin{align}
            \textnormal{dim}(W^{\textnormal{u}}_\textnormal{loc}) = \begin{cases}
                N_{-}, &\textnormal{ for } X_I > \frac{N_{+}}{N}k A - \alpha^{-1}\\
                N_{-} + 1, &\textnormal{ for } X_I \leq \frac{N_{+}}{N}k A - \alpha^{-1}.
            \end{cases} \nonumber
        \end{align}
\end{itemize}
}

\noindent
\textit{Proof.} In the case of $N_{+} = 0$, we obtain from both reduced eigenvalue problems (\ref{eq: off part of reduced remaining case}), (\ref{eq: reduced eigenvalue problem OFF, N+ = 0}) the eigenvalues $\lambda_1 = - \gamma$ and $\lambda_2 = x_{I}$ of which only $\lambda_2$ is unstable for $\omega \geq \frac{1 - \delta}{k}$. Clearly, since there are two distinct eigenvalues for each of the reduced system of size 2, each eigenvalue has a geometric multiplicity of one. Say these eigenvectors of $\lambda_2$ are $\mathbf{b}_2$ and $\mathbf{z}_2$ of the reduced eigenvalue problems (\ref{eq: off part of reduced remaining case}), (\ref{eq: reduced eigenvalue problem OFF, N+ = 0}) respectively, then the eigenvectors of the original eigenvalue problem are  $(\mathbf{w}_0 \otimes \mathbf{b}_2), (\mathbf{w}_1 \otimes \mathbf{z}_2), \ldots, (\mathbf{w}_{N_{-} - 1} \otimes \mathbf{z}_2) $. Therefore, 
\begin{align}
    \text{dim}(W^{\text{u}}_{\text{loc}}) = N_{-} = N. \nonumber
\end{align}
For the remaining case $0 < N_{+} \leq N$, we use the following property of the solutions $\lambda_3$ and $\lambda_4$ of the polynomial \eqref{eq:Polynom}:
\begin{itemize}
    \item $\lambda_3<0$ and $\lambda_4<0$ for $X_I > \frac{N_{+}}{N}k A - \frac{1}{\alpha}$;
    \item $\lambda_3>0$ and $\lambda_4<0$ for $X_I \le \frac{N_{+}}{N}k A - \frac{1}{\alpha}$.
\end{itemize}

Let $\mathbf{a}_3$ be the eigenvector of $\lambda_3$. Then we have
\begin{itemize}
    \item[]Case $0 < N_{+} < N$ and $X_I \leq \frac{N_{+}}{N}k A - \frac{1}{\alpha}$:
    \begin{align}
        \text{dim}(W^{\text{u}}_{\text{loc}}) 
        = \text{dim}(\{ (\mathbf{v}_0 \otimes \mathbf{a}_3), (\mathbf{w}_0 \otimes \mathbf{b}_2), (\mathbf{w}_1 \otimes \mathbf{z}_2), \ldots, (\mathbf{w}_{N_{-} - 1} \otimes \mathbf{z}_2) \}) 
       = N_{-} + 1 \nonumber 
    \end{align}
    \item[]Case $0 < N_{+} < N$ and $X_I > \frac{N_{+}}{N}k A - \frac{1}{\alpha}$:
    \begin{align}
        \text{dim}(W^{\text{u}}_{\text{loc}}) = \text{dim}(\{ (\mathbf{w}_0 \otimes \mathbf{b}_2), (\mathbf{w}_1 \otimes \mathbf{z}_2), \ldots, (\mathbf{w}_{N_{-} - 1} \otimes \mathbf{z}_2) \})
        = N_{-} \nonumber 
    \end{align}
    \item[]Case $N_{+} = N$ and $X_I \leq \frac{N_{+}}{N}k A - \frac{1}{\alpha}$:
    \begin{align}
        \text{dim}(W^{\text{u}}_{\text{loc}}) = \text{dim}(\{ (\mathbf{v}_0 \otimes \mathbf{a}_3) \})  = 1 = N_{-} + 1 \nonumber
    \end{align}
    \item[]Case $N_{+} = N$ and $X_I > \frac{N_{+}}{N}k A - \frac{1}{\alpha}$:
    \begin{align}
        \text{dim}(W^{\text{u}}_{\text{loc}}) = 0  \nonumber
    \end{align}
\end{itemize}
where, besides the result for $N_{+}=0$, we additionally used that 
$\mathbf{v}_0$ and $\{\mathbf{w}_m\}_{m=0, \ldots, N_{-} - 1}$ are linearly independent.

\noindent
\textit{Proof is complete.}

Lemma 6 immediately implies that $\textnormal{dim}(W^{\textnormal{u}}_{\textnormal{loc}}(B_{000})) + \textnormal{dim}(W^{\textnormal{s}}_{\textnormal{loc}}(B_J)) = 2N+1$, where $J$ is one of the branches with a single laser on, provided the parameter $\omega$ is sufficiently large (beyond the branching points). The same holds for the transition between neighboring branches where $N_-$ changes by one, provided $X_I > \frac{N_{+}}{N}k A - \alpha^{-1}$, which means that the branch is considered beyond the fold point. 

\begin{figure}[h]
    \includegraphics[width=\textwidth]{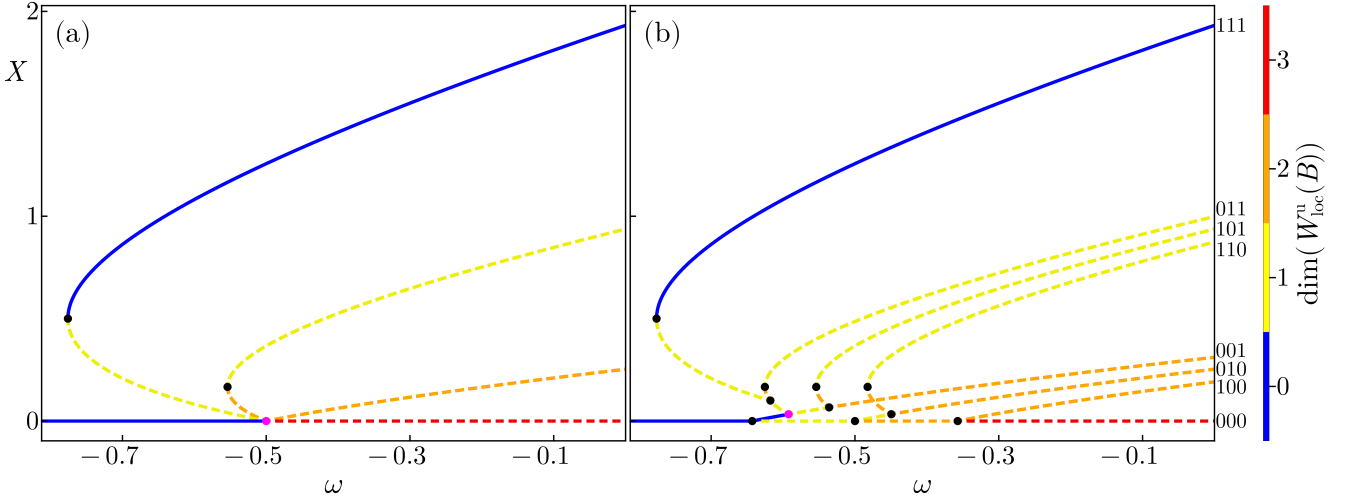} 
\caption{Dimensions of unstable manifolds $\dim(W^\textnormal{u}_{\textnormal{loc}})$ along the branches of the critical manifold. The dimensions correspond to the number of unstable eigenvalues of the fast subsystem for fixed $\omega$; their number changes at the transcritical and fold points (black dots). The transcritical points which define the beginning of the examined interval of heteroclinic connections $\omega \in (\omega_b, 0]$ are shown in magenta (see Fig.~\ref{fig: heteroclinic connections with min genericity condition}). Parameter values are the same as in Fig.~\ref{fig:symmetric_cm_with_dynamics} for (a) and as in Fig.~3 of the main manuscript. 
\label{fig: cm with dim of unstable manifold} 
} 
\end{figure}

The results of Lemma 6 are illustrated in Fig.~\ref{fig: cm with dim of unstable manifold}(a) for the case of three lasers. 
From the results for the symmetric case, the asymmetric case with weakly inhomogeneous $\delta_i$ can be deduced using continuation arguments and the fact that hyperbolic equilibria are robust to small parameter changes, see Fig.~\ref{fig: cm with dim of unstable manifold}(b). This means that Lemma 6 can be applied to a weakly non-symmetric case.

\begin{figure}[h]
\includegraphics[width=\textwidth]{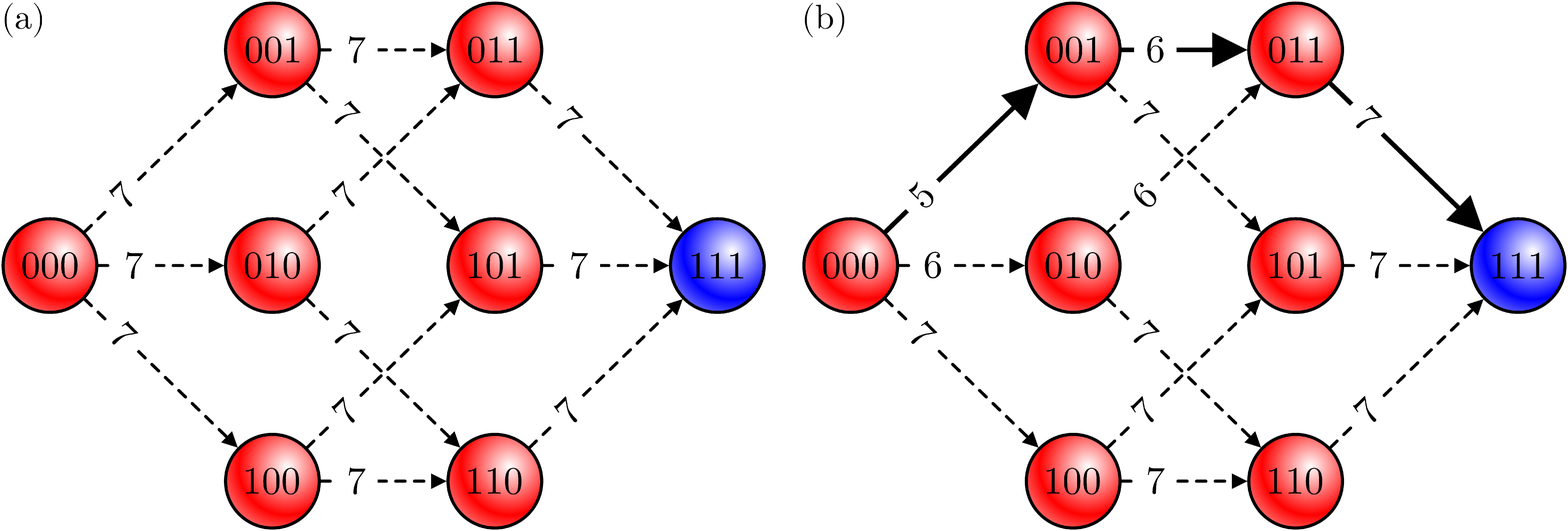}

\caption{Graph of possible heteroclinic connections. Edges are annotated by the minimal value of the sum $\textnormal{dim}(W^{\textnormal{u}}_{\textnormal{loc}}(B_{I})) + \textnormal{dim}(W^{\textnormal{s}}_{\textnormal{loc}}(B_J))$  for the existing connections $B_{I}\to B_J$, see eq. (\ref{genericity_condition}) for $\omega \in (\omega_b,0]$, where $\omega_b = \frac{1 - \delta}{k}$ and all $\delta_i$ are the same for the symmetric case, $\omega_b = \frac{1 - \delta_2}{k} - A \ln \left( 1 + \alpha X_{001} \right)$ for the nonsymmetric case. 
}

\label{fig: heteroclinic connections with min genericity condition}

\end{figure}

Figure~\ref{fig: heteroclinic connections with min genericity condition} illustrates the graphs of possible heteroclinic connections with the edges annotated by the minimal value of the sum of the dimensions $\textnormal{dim}(W^{\textnormal{u}}_{\textnormal{loc}}(B_{I})) + \textnormal{dim}(W^{\textnormal{s}}_{\textnormal{loc}}(B_J))$. Before proceeding, for the non-symmetric case, we need to further specify what is meant by "graph of possible heteroclinic connections". For this, one can convince oneself by comparing Figs.~\ref{fig: heteroclinic network transitions illustration}(b), (c), (e), (f), (g), (i), (j) and Fig.~\ref{fig: heteroclinic connections with min genericity condition}(b) that the graphs for different values of $\omega$ (not considering fold and branching points) are subgraphs of Fig.~\ref{fig: heteroclinic connections with min genericity condition}. 
For the symmetric case there is no need for these considerations as the graph of possible connections and the actual graph of connections coincide.
Consequently Figure~\ref{fig: heteroclinic connections with min genericity condition}(a) shows the case that is given by Lemma~6 for the symmetric system beyond the branching point $\omega > \omega_b = \frac{1-\delta}{k}$. Note that in the asymmetric case in Fig.~\ref{fig: heteroclinic connections with min genericity condition}(b) the condition (\ref{genericity_condition}) is not fulfilled, but the heteroclinic connections do exist due to the existence of invariant subspaces.
Within these invariant subspaces, the corresponding heteroclinic orbit connects a saddle with one unstable direction to an attractor. 
For example, the connection  $B_{000}\rightarrow B_{001}$ is realized within a 4-dimensional invariant subspace defined by $x_1 = x_2 = 0$. Hence, the condition (\ref{genericity_condition}) holds for $N=4$.

\section{A.IV. Jump points}
\label{sec: jump points}

Let us first consider the jump from the branch $B_{000}$ in the case of three lasers. The motion during this jump is illustrated in Fig.~4 of the main manuscript, see the part AC:
\begin{itemize}
    \item The part AB: convergence to the branch $B_{000}$ with the rate corresponding to the maximal (least stable) eigenvalue of the corresponding equilibrium of the fast subsystem; we call it $\mu_1^{000}(\omega)$, where we have explicitly written the dependence of this eigenvalue on $\omega$. 
    \item BC: repulsion from the branch $B_{000}$. We assume that there the eigenvalues do not change their order, and this repulsion takes place with the same rate $\mu_1^{000}(\omega)$, where $\mu_1^{000}(\omega)$ changes its sign to positive at $\omega_B$ in point B.
\end{itemize}

Thus, the small $\Delta$ distance to the $B_{000}$ branch can be approximated by the linearized dynamics within the leading manifold $W^l(B_{000})$
\begin{align}
    \frac{\text{d}\Delta}{\text{d}t}  \simeq \mu_1^{000}(\omega(t)) \Delta(t). \label{eq: jump point ansatz}
\end{align}
As a side note, we have used here that the branch $B_{000}$ does not change with $\omega$. The latter equation can be rewritten with respect to $\omega$ instead of time:
\begin{align}
    \frac{\text{d}\Delta}{\text{d}t} &= \frac{\text{d} \Delta}{\text{d} \omega} \frac{\text{d} \omega}{\text{d}t} = \frac{\text{d} \Delta}{\text{d} \omega} \left[-\varepsilon (\omega + f(X_{000})) \right], \nonumber \\
    \Rightarrow \frac{\text{d} \Delta}{\text{d} \omega} &= - \frac{1}{\varepsilon (\omega + f(X_{000}))} \mu_1^{000}(\omega) \Delta(\omega) , \label{eq: jump point ansatz in omega}
\end{align}
and the subsequent solution of the differential equation in (\ref{eq: jump point ansatz in omega}) gives
\begin{align}
    \Delta(\omega_{C}) = \Delta(\omega_A) \exp \left[ - \frac{1}{\varepsilon} \int_{\omega_A}^{\omega_{C}} \frac{\mu_1^{000}(\omega)}{\omega + f(X_{000})} \, \text{d}\omega \right], \nonumber
\end{align}
where $\omega_A$ is the drop point, i.e. the point where the dynamics return to the $B_{000}$ branch, $X_{000}$ is the mean field at the branch $B_{000}$, and $\omega_{C}$ denotes the $\omega$-value of the first jump. Then we assume that the jump point is given 
\begin{align}
    \Delta(\omega_{C}) \approx \Delta(\omega_A), \nonumber
\end{align}
resulting in the final jump point condition
\begin{align}
    \int_{\omega_A}^{\omega_{C}} \frac{\mu_1^{000}(\omega)}{\omega + f(X_{000})} \, \text{d} \omega = 0. \label{eq: jump condition first jump}
\end{align}

Following similar reasoning, the second jump from the branch $B_{001}$ to $B_{011}$ (see point E in Fig.~4 of the main manuscript) is described by the whole trajectory part AE:
\begin{itemize}
    \item AB: Convergence to the leading manifold $W^l(B_{000})$ corresponding to the second least stable direction along $B_{000}$. This convergence is governed by the second largest eigenvalue of the corresponding equilibrium of the fast subsystem; we call it $\mu_2^{000}(\omega)$.
    \item BC: In part BC, the leading manifold is still exponentially stable with the convergence rate given by $\mu_2^{000}(\omega)$.
    \item CD: At point C, the orbit jumps to the vicinity of the branch $B_{001}$. This happens on the fast timescale, so the contribution to the distance to $W^l(B_{000})$ is negligible. 
    \item DE: Starting from the point D, the repulsion from the branch $B_{001}$ (and from $W^l(B_{000})$) takes place at the exponential rate governed by the most unstable eigenvalue $\mu_1^{001}$ of the branch $B_{001}$. 
\end{itemize}
Summing up the convergence and repulsion processes described above, we obtain the following condition for the coordinate $\omega_{E}$ of the jump point E:
\begin{align}
    \int_{\omega_A}^{\omega_{C}} \frac{\mu_2^{000}(\omega)}{\omega + f(X_{000}))} \, \text{d} \omega + \int_{\omega_{C}}^{\omega_{E}} \frac{\mu_1^{001}(\omega)}{\omega + f(X_{001}))} \, \text{d} \omega = 0 . \label{eq: jump condition second jump}
\end{align}
Note that we have neglected the $\omega$ dependence of the $B_{001}$ branch, which will probably lead to an error even if $\varepsilon\to 0$. However, the numerical test in Fig.~3 shows that \eqref{eq: jump condition second jump} is already a decent approximation, and we decided not to go into more quantitative details, since the qualitative picture is clear. 

Using similar arguments, we obtain an approximation for the third jump point:
\begin{equation}
    \label{eq:jump3}
    \int_{\omega_A}^{\omega_{C}} \frac{\mu_3^{000}(\omega)}{\omega + f(X_{000})} \, \text{d} \omega + \int_{\omega_{C}}^{\omega_{E}} \frac{\mu_2^{001}(\omega)}{\omega + f(X_{001})} \, \text{d} \omega \nonumber + \int_{\omega_{E}}^{\omega_{G}} \frac{\mu_1^{011}(\omega)}{\omega + f(X_{011})} \, \text{d} \omega = 0. 
\end{equation}
Numerical results for our example using the obtained approximations \eqref{eq: jump condition first jump}, \eqref{eq: jump condition second jump}, \eqref{eq:jump3} are shown in Fig.~3 of the main manuscript.

\end{document}